\documentclass[%
 reprint,
 amsmath,amssymb,
 aps,
pra,
]{revtex4-2}

\usepackage[dvipsnames, svgnames, x11names]{xcolor}
\usepackage{graphicx}
\usepackage{dcolumn}
\usepackage{bm}
\usepackage{amsmath}
\usepackage{multirow}
\usepackage{booktabs}
\usepackage{amsfonts}
\usepackage{makecell}
\usepackage{float}
\usepackage{diagbox}
\usepackage{tabularx}
\usepackage{tabu}
\usepackage{natbib}
\definecolor{deepgreen}{rgb}{0, 0.5, 0}
\begin{document}

\preprint{APS/123-QED}

\title{Active robustness against the detuning-error for Rydberg quantum gates}
\thanks{jqian1982@gmail.com}%

\author{Qing-Ling Hou$^{1}$}
\author{Han Wang$^{1,2}$}
\author{Jing Qian$^{1,3}$} 

\affiliation{$^{1}$State Key Laboratory of Precision Spectroscopy, Department of Physics, School of Physics and Electronic Science, East China Normal University, Shanghai, 200062, China
}
\affiliation{$^{2}$Hong Kong University of Science and Technology, Guangzhou, 511453, China}
\affiliation{$^{3}$Shanghai Branch, Hefei National Laboratory, Shanghai 201315, China}




\date{\today}

\begin{abstract}
Error suppression to the experimental imperfections is a central challenge for useful quantum computing. Recent studies have shown the advantages of using single-modulated pulses based on optimal control which can realize high-fidelity two-qubit gates in neutral-atom arrays. However, typical optimization only minimizes the ideal gate error in the absence of any decay,
which allows the gate to be passively influenced by all error sources leading to an exponential increase of the insensitivity when error becomes larger. In the present work, we 
propose the realization of two-qubit $C_Z$ gates with \textit{active} robustness against two-photon detuning errors. Our method depends on a modified cost function in numerical optimization for shaping gate pulses, which minimizes, not only the ideal gate error but also the fluctuations of gate infidelity over a wide error range. We introduce a family of Rydberg blockade gates with
{\it active} robustness towards the impacts of versatile noise sources such as Doppler dephasing and ac Stark shifts.
 The resulting gates with robust pulses can significantly increase the insensitivity to any type of errors acting on the two-photon detuning, benefiting from a relaxed requirement of colder atomic temperatures or more stable lasers for current experimental technology.
\end{abstract}

\maketitle


\maketitle

\section{Introduction}

The ability to achieve high-fidelity two-qubit quantum gates is a key task for large-scale quantum computing with trapped neutral atoms \textcolor{black}{\cite{Nature.453.1016(2008),Nature.464.45(2010),RevModPhys.82.2313,QuantumSci.Technol7.023002(2022),shao2024rydberg}}, while noises from versatile experimental imperfections will fundamentally limit the achievable gate fidelity \cite{Nature.626.58(2024)}, leading to a huge gap to other state-of-the-art platforms such as superconducting qubits, trapped ions, quantum dots and nuclear magnetic resonance systems \textcolor{black}{\cite{PhysRevLett.125.240503,PhysRevApplied.14.054062,Nature.555.75-78(2018),PhysRevLett.119.140503,PhysRevApplied.12.024024}}. Recent work has demonstrated that the execution of modulated pulses based on optimal control can realize high-fidelity entangling operations \cite{PhysRevA.105.042430}. And to date the best reported two-qubit gate fidelity using optimal control and atomic dark state reaches $\mathcal{F} = 0.995$ \textcolor{black}{\cite{2023Nature}}. The optimal control technique enables a smooth modulation of the laser amplitude or phase in time without the requirement for individual addressability of atoms \textcolor{black}{\cite{PhysRevApplied.13.024059,PhysRevApplied.17.024014,NewJ.Phys.25(2023)123007,PhysRevLett.131.170601,Science.376.1209-1215(2022),PhysRevA.109.022613}}. This outperforms the original $\pi$-gap-$\pi$ method \textcolor{black}{\cite{PhysRevLett.85.2208,PhysRevA.88.062337,PhysRevApplied.15.054020}} by suppressing severe decoherence between the ground and Rydberg states \cite{PhysRevApplied.15.054020,PhysRevApplied.11.044035,QuantumSci.Technol.4.015011(2018)}.

The typical optimization focuses on minimizing the ideal gate infidelity in the absence of any noise. {\it e.g.} A time-optimal gate contains the merit of minimizing the gate duration together with a smaller time-spent in the Rydberg state which makes the gate intrinsically robust to the decay errors \textcolor{black}{\cite{PhysRevResearch.4.033019}}. Whereas, other technical errors will passively affect the gate performance arising the necessity of subsequent error correction ways which is regardless of the system
\textcolor{black}{\cite{PhysRevLett.129.030501,PhysRevX.11.041058,Nature.598.281(2021),Nature.605.675(2022),Nature.606.884(2022),Nature.605.669(2022)}}.
Recently, performing low-error quantum gates with 
noise-robust optimization has become a new exciting frontier. To alleviate the insensitivity of gates to type of errors that occur, {\it e.g.} constructing a specific cost function that possesses enhanced robustness with respect to some errors, is more attractive \textcolor{black}{\cite{PhysRevResearch.4.023155,NewJ.Phys.26.013024(2024),PhysRevApplied.21.044012}}. Via utilizing quantum optimal control with GRAPE algorithm \cite{Quantum.6.712(2022)} 
a family of Rydberg blockade gates with significantly improved performance is presented that can be simultaneously robust against the amplitude inhomogeneity and the Doppler shifts {\cite{PRXQuantum.4.020336}}. Although these works may open unique opportunities for fault-tolerant quantum computing with atomic qubits, preserving an \textit{active} robustness of systems to type of errors for Rydberg gates is still a big challenge.

\begin{figure*}
    \centering
    \includegraphics[scale=0.56]{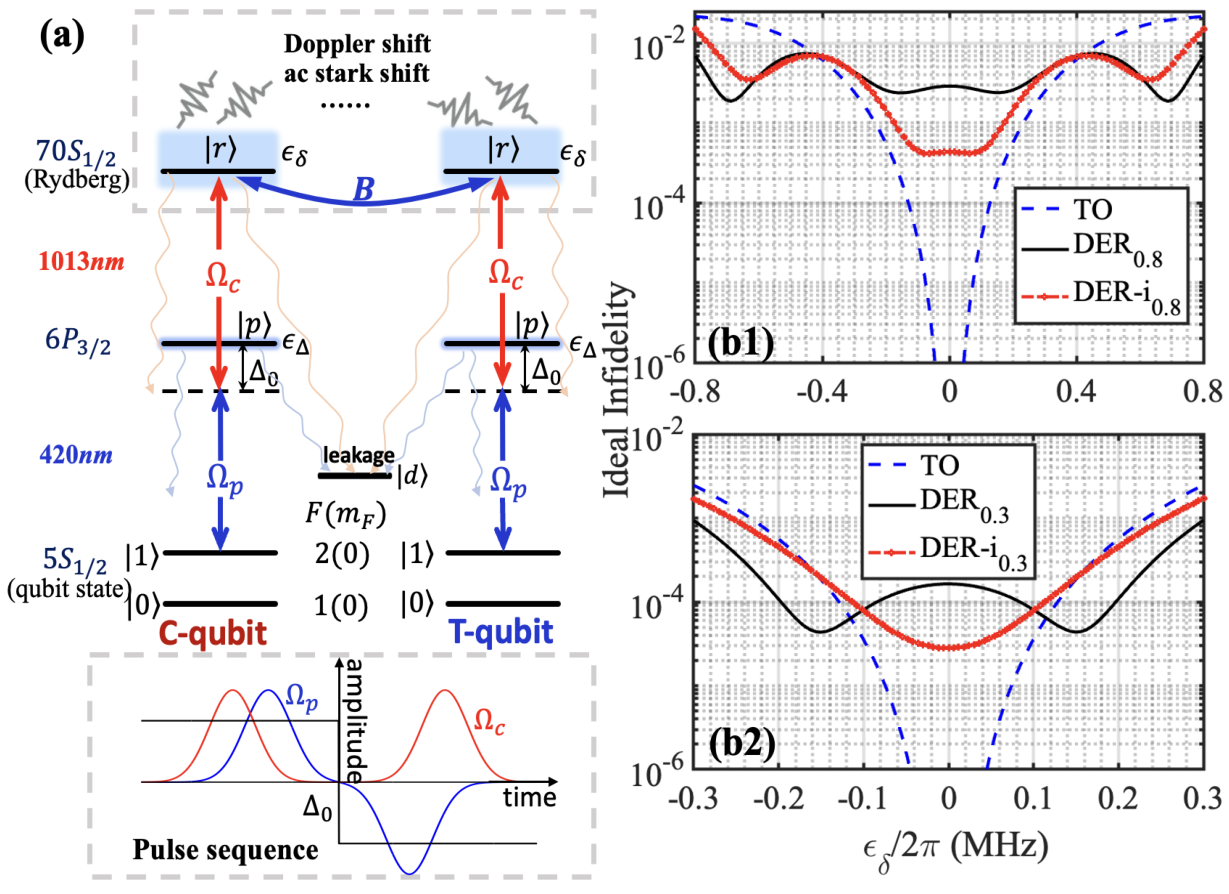}
    \caption{Two-qubit $C_Z$ gate protocol. (a) Atom-level diagram. Qubits are encoded in two hyperfine clock states $|0\rangle$ and $|1\rangle$ of $^{87}\rm{Rb}$ atoms. Atomic excitation  to the Rydberg state $|r\rangle$ is carried out by a two-photon transition via a largely-detuned intermediate state $|p\rangle$(detuned by $\Delta_0$), driven by 420 nm and 1013 nm lasers.
    $|d\rangle$ is a leakage state that collects most of the spontaneous decays from $|p\rangle$ and $|r\rangle$ except returning to $|0\rangle$ and $|1\rangle$. The branching radios used for different scattering rates are: $b_{r\to0}=0.059$, $b_{r\to1}=0.055$, $b_{r\to d}=0.886$, and $b_{p\to0}=0.1354$, $b_{p\to1}=0.2504$, $b_{p\to d}=0.6142$, together with a $1/\gamma_r=375$ $\mu$s lifetime for $|r\rangle$ and a $1/\gamma_p=0.118$ $\mu$s lifetime for $|p\rangle$, chosen from Ref. \cite{PhysRevA.85.033817}. The van der Waals interaction leads to a large energy shift $B$ if both atoms are in $|r\rangle$ preventing their simultaneous excitation in the Rydberg blockade limit. $\epsilon_\Delta$ and $\epsilon_\delta$ represent unknown detuning errors respectively to the intermediate and Rydberg states possibly from versatile experimental imperfections. Bottom of (a) presents typical profiles of the adiabatic pulse sequences $\Omega_p(t),\Omega_c(t)$ and the non-fluctuated intermediate detuning $\Delta_0$. Note that $\Delta_0$ is switched between two STIRAP pulses. \textcolor{black}{(b1-b2) The ideal gate infidelity ($\gamma_{p,r}=0$) as a function of the two-photon detuning error $\epsilon_\delta$ for different DER and DER-i pulses, where the optimization parameter ranges are $\epsilon_0/2\pi=(0.8,0.3)$ and the TO pulse is same.}
    }
    \label{Fig1:model}
\end{figure*}

In this work, we present the realization of two-qubit Rydberg $C_Z$ gates with {\it active} robustness to all error sources associated with the variation in two-photon detuning \cite{PhysRevA.91.012337,PRXQuantum.4.020335}. To implement the noise-robust optimization, we modify the cost function in the numerical Genetic algorithm, which minimizes, not only the ideal gate error but also the infidelity fluctuations over a wide error range, enabling the optimal pulse to be {\it actively} robust against the variation of detuning errors.
In addition, we adopt a realistic model by considering a native two-photon transition using adiabatic double-STIRAP pulse \cite{PhysRevA.101.062309} which has shown a better robustness to the variation of experimental parameters such as laser amplitude and pulse shape \textcolor{black}{\cite{NewJ.Phys.19(2017)093016,PhysRevA.102.023715,PhysRevA.107.012605}}. To characterize the resulting two-qubit $C_Z$ gates, we comparably exploit the results from typical and noise-robust optimizations. The latter can produce experimentally realizable pulses that are more robust to any deviation of two-photon detuning, caused by Doppler dephasing {\cite{PhysRevA.103.012601}} and ac Stark shift \cite{PhysRevLett.98.203005,PhysRevA.92.022336}. Our approach can be easily generalized to other type of errors paving the avenue for realizing noise-robust quantum gates in future experiments.

Finally we admit that there are two main weakness of this protocol. First, the modified cost function contains two competing terms that means the improvement of robustness should be at the expense of low gate fidelity at small detuning deviations (see Appendix B). For a noise-robust pulse ({\it e.g.} the DER pulse), we only achieve the gate with an ideal fidelity of $\mathcal{F}_{der} \approx 0.9971$, depending on a wide range of optimization for enhancing the robustness. While a higher value $\mathcal{F}_{der-i}>0.9999$ is possible by utilizing the fine optimization of the DER-i pulse.
Second, the proposed gates will be passively affected by other error sources not associated with the two-photon detuning. As compared to a typical-optimal pulse, our pulses are not explicitly robust to those errors (see Appendix C). Especially, due to the use of an adiabatic pulse sequence that elongates the gate duration, our scheme is more susceptible to the fluctuation in laser phase \textcolor{black}{\cite{PhysRevA.107.042611,PhysRevApplied.20.014014}}.

\section{Model description}

 Consider a level scheme for two-qubit $C_Z$ gates, as shown in Fig.\ref{Fig1:model}(a). Each atom is modeled as a native four-level system with long-lived hyperfine qubit states $|0\rangle,|1\rangle$, an intermediate state $|p\rangle$ and an uppermost Rydberg state $|r\rangle$ \textcolor{black}{\cite{PhysRevA.96.043411}}. The coherent excitation from $|1\rangle$ to $|r\rangle$ is mediated by a resonant two-photon transition with laser Rabi frequencies $\Omega_p,\Omega_c$, largely detuned by $\Delta_0$ from the intermediate state $|p\rangle$, {\it i.e.} $|\Delta_0|\gg \Omega_p,\Omega_c$. We assume a strong Rydberg blockade condition where two atoms trapped in two optical tweezers \textcolor{black}{\cite{PhysRevA.106.022611}} with a distance smaller than the blockade radius,  can bring a strong energy shift $B$ to prevent simultaneous excitation of both atoms \textcolor{black}{\cite{NaturePhysics.5.110(2009),Naturephysics1183,PhysRevLett.110.263201}}. 

The total Hamiltonian governing the dynamics of the system, reads 
\begin{equation}
{\hat{H}}={\hat{H}_c}\otimes \hat{I} + \hat{I} \otimes {\hat{H}_t} +B|rr\rangle\langle rr|
\end{equation}
with
\begin{eqnarray}
    {\hat{H}_{j\in(c,t)}} 
    &=&\frac{\Omega_p}{2}|p\rangle_j \langle 1|+\frac{\Omega_c}{2}|r\rangle_j \langle p|+\text{H.c.}- \epsilon_\delta|r\rangle_j \langle r| \nonumber \\
    & -&(\Delta_0+\epsilon_\Delta)|p\rangle_j \langle p|\label{Hj}
\end{eqnarray}
 for single control or target qubit. Here, $\epsilon_\Delta$ and $\epsilon_\delta$ treats as the unknown yet constant detuning error over the gate duration, possibly caused by versatile noise sources such as atomic velocity and laser fluctuations.
 We point out that, the intermediate detuning error $\epsilon_\Delta$ is less important because $|\Delta_0|\gg \epsilon_\Delta$ arising a negligible impact $10^{-10}\sim 10^{-6}$ on fidelity (see Appendix C) \textcolor{black}{\cite{PhysRevA.105.043715}}. Nevertheless, the laser frequency is resonant with the two-photon transition frequency between $|1\rangle$ and $|r\rangle$, making the role of two-photon detuning error $\epsilon_\delta$ more crucial in the scheme.

\section{Active-robustness two-qubit $C_Z$ Gates}


\subsection{Double STIRAP pulse sequences}

We proceed by describing the implementation of specific pulse sequences that result in a $\pi$-phase($C_Z$) gate. The requirement for a universal controlled-phase gate is that the laser pulses $\Omega_{p,c}(t)$ and the detuning $\Delta_0$(in the absence of any error) are applied such that states $|01\rangle$, $|10\rangle$, $|11\rangle$ return to their ground states with phases $\phi_{01}$, $\phi_{10}$, $\phi_{11}$, in which $\phi_{01}=\phi_{10}$ by symmetry. If
$ \phi_{11} =   2\phi_{01} -\pi $ is satisfied it is so-called a $C_Z$ gate \textcolor{black}{\cite{PhysRevLett.123.170503}}. To realize this gate, we resort to the protocol by using adiabatic double STIRAP pulse, which in principle provides strongly-suppressed insensitivity to the fluctuations in laser amplitude \textcolor{black}{\cite{PhysRevA.89.013831}}. The adiabatic pulses are given by
\begin{eqnarray}
\Omega_{p}(t) &=& \Omega_{p}^{\max}(e^{-\frac{(t+t_1)^2}{2\omega_1^2}} 
 - e^{-\frac{(t-t_1)^2}{2\omega_1^2}})\label{opa} \\
\Omega_{c}(t) &=& \Omega_{c}^{\max}(e^{-\frac{(t+t_2)^2}{2\omega_2^2}} +  e^{-\frac{(t-t_2)^2}{2\omega_2^2}}) \label{opb}
\end{eqnarray}
with $\Omega_{p,c}^{\max}$, $t_{1,2}$, $\omega_{1,2}$ the peak amplitude, the pulse centers and the pulse widths, respectively.
Note that, in addition to the phase change of $\Omega_p(t)$ by $\pi$ in the second STIRAP sequence, the sign of intermediate-state detuning $\Delta_0$ is also reversed in the middle of the gate \cite{PhysRevA.88.010303}, as shown in Fig.\ref{Fig1:model}a (bottom).

To understand the reason of this pulse (detuning) reversion, we demonstrate the physics associated with a single three-level system $\{|1\rangle,|p\rangle,|r\rangle\}$ at which the Hamiltonian is readily given by (equivalent to Eq.\ref{Hj} without detuning errors $\epsilon_\Delta,\epsilon_\delta$)
\begin{equation}
{\hat{H}} = 
	\begin{bmatrix}
	0& \frac{\Omega_p}{2}& 0 \\
	  \frac{\Omega_p}{2} & -\Delta_0& \frac{\Omega_c}{2} \\
	  0 & \frac{\Omega_c}{2}  & 0 
	 \end{bmatrix}
  \label{Hom}
\end{equation}

At large intermediate detuning $|\Delta_0|\gg \Omega_{p,c}^{\max}$, the system is conveniently described by
one adiabatic dark state $|d\rangle$ and two bright states $|+\rangle,|-\rangle$, which are
\begin{eqnarray}
|d\rangle &=& -\frac{1}{\sqrt{1+\alpha^2}}|1\rangle + \frac{\alpha}{\sqrt{1+\alpha^2}}|r\rangle \nonumber
\\
|+\rangle &=& \frac{\alpha}{\sqrt{1+\alpha^2}}|1\rangle+\frac{\sqrt{1+\alpha^2}\Omega_{c}}{2\Delta_0}|p\rangle + \frac{1}{\sqrt{1+\alpha^2}}|r\rangle \nonumber
 \\
|-\rangle &=& -\frac{\alpha\Omega_{c}}{2\Delta_0}|1\rangle + |p\rangle - \frac{\Omega_{c}}{2\Delta_0}|r\rangle \label{dd}
\end{eqnarray}
with the corresponding eigenvalues $\lambda_{d}=0$, $\lambda_{+}\approx\frac{(1+\alpha^{2})}{4\Delta_0}$, $\lambda_{-}\approx-\Delta_0$ to the leading order in $\frac{(1+\alpha)^2\Omega_{c}^{2}}{\Delta_0^2}$, where $\alpha=\Omega_p/\Omega_c$. Reversing the sign of intermediate detuning $\Delta_0\to-\Delta_0$ can make the accumulated dynamical phase vanish due to $\int_{-T_g/2}^{T_g/2}\lambda_{d,+,-} dt=0$ ($T_g$ is the gate duration). Thereby an additional geometric phase 
 \textcolor{black}{\cite{berry1984quantal}} arises owing to the relative phase switching between $\Omega_p$ and $\Omega_c$ 
which gives an extra $\pi$ phase difference between $|1\rangle$ and $|r\rangle$ because of $\alpha\rightarrow-\alpha$ in the second STIRAP sequence \textcolor{black}{\cite{PhysRevA.75.062302}}. Such a pulse inversion can result in the state evolution obeying $|01\rangle\to -|0r\rangle\to-|01\rangle$, $|10\rangle\to -|r0\rangle\to-|10\rangle$ so $\phi_{01}=\phi_{10}=\pi$. Besides, the $|11\rangle$ state evolves within the Rydberg blockade regime as an effective two-level system by following $|11\rangle\rightarrow-\frac{|1r\rangle+|r1\rangle}{\sqrt{2}}\rightarrow-|11\rangle$ and also accumulates $\phi_{11}=\pi$, which finally realizes the $C_Z$ gate for $\phi_{11} = -\pi + \phi_{01} + \phi_{10}$. \textcolor{black}{The derivation for the effective two-level couplings and numerical verification of the gate are presented in Appendix A}.

\begin{table*}
\caption{Optimized gate parameters for the pulse centers $t_1,t_2$ and pulse widths $\omega_1,\omega_2$. The first pair of DER-i pulses (row 6 and row 7) is associated with Gaussian-type weights while the second pair (row 8 and row 9) is obtained by choosing a set of Uniform-type weights.
The total gate duration is solved by $T_{g}=2(t_{2}+3\omega_{2})$. The last column presents the ideal gate fidelity at $\epsilon_\delta=0$.}\label{tab 1}
\setlength{\tabcolsep}{13.0pt}
\renewcommand\arraystretch{1.5}
\begin{tabular}{c|c|c|c|c|c|c|c} 
 \hline\hline 
 \multicolumn{2}{c|}{Optimization case} & \multicolumn{4}{c|}{Pulse parameters ($\mu$s)}& \multicolumn{2}{c}{Gate performance}  \\
 \hline
 Robust pulse  & $\epsilon_{0}/2\pi$ (MHz) & $t_{1}$ & $t_{2}$ & $\omega_{1}$ &  $\omega_{2}$ & $T_{g}$ &$\mathcal{F}$($\epsilon_\delta=0$)\\
 \hline
 TO  & $\times$ & 0.4277 & 0.6697 & 0.1584 & 0.1584 & 2.2898 & 0.9999997\\
 \hline
 $DER$  &0.8& 0.6664 &0.9259 &  0.1666 &0.1665 & 2.8509& 0.9970983 \\
   \hline
$DER$& 0.3 & 0.2863 & 0.5332 & 0.1604 &  0.1586 & 2.0180 &0.9998366\\
 \hline
 $DER-i$ & 0.8 & 0.7261 &0.9737 &0.1619 &  0.1601 & 2.9080 & 0.9995629\\
 \hline
  $DER-i$& 0.3 & 0.2874 & 0.5310 & 0.1598 &  0.1588 & 2.0130 &0.9999807\\
 \hline\hline
   \textcolor{black}{$DER-i$}& \textcolor{black}{0.8} & \textcolor{black}{0.7285} & \textcolor{black}{0.9788} & \textcolor{black}{0.1619} &  \textcolor{black}{0.1631} & \textcolor{black}{2.9363} & \textcolor{black}{0.9991733}\\
    \hline
  \textcolor{black}{$DER-i$}& \textcolor{black}{0.3} & \textcolor{black}{0.2722} & \textcolor{black}{0.5196} & \textcolor{black}{0.1600} &  \textcolor{black}{0.1591} & \textcolor{black}{1.9938} & \textcolor{black}{0.9998411}\\
 \hline\hline
\end{tabular}
\label{tableI}
\end{table*}

\subsection{Numerical calculation method}

The $C_Z$ gate performance is quantified by the gate fidelity.  
For any input state $|\Psi_{j}\rangle\in\{|00\rangle,|01\rangle,|10\rangle,|11\rangle\}$( $j=1,...,4$), the initial matrix can be expressed as 
\begin{equation}
    \hat{\rho}_{in}=|\Psi_j\rangle \langle \Psi_j|
\end{equation}
 and its dynamical evolution $\hat{\rho}(t)$ can be analyzed by numerical integration of the two-atom master equation in the Lindblad form \cite{PhysRevA.95.012708}
\begin{equation}
    \frac{\partial \hat{\rho}}{\partial t} = -i [\hat{H},\hat{\rho}]+ \hat{\mathcal{L}}[\hat{\rho}] \label{rho}
\end{equation}
where the Lindblad decay term is
\begin{equation}
    {\hat{\mathcal{L}}}[{\hat{\rho}}] = \sum_{l\in\{c,t\}}\sum_{j \in\{p,r\}}\sum_{i\in\{0,1,d\}}{ {\hat{\mathcal{L}}}^{l}}_{ij}{\hat{\rho}}{ {\hat{\mathcal{L}}}^
{l^{\dagger}}}_{ij}-\frac{1}{2}\{{ {\hat{\mathcal{L}}}^{l^{\dagger}}}_{ij}{ {\hat{\mathcal{L}}}^{l}}_{ij},{\hat{\rho}}\}
\end{equation}
with jump operators ${ {\hat{\mathcal{L}}}^{l}}_{ij} =\sqrt{\gamma_{j}b_{j\to i}} |i\rangle\langle j| $ corresponding to the spontaneous decay from state $|j\rangle$ with rate $\gamma_j$ and branching ratio $b_{j\to i}$ \cite{PhysRevA.89.030301}. The uncoupled state $|d\rangle$ represents all hyperfine ground states outside the qubit basis and all population leakage into $|d\rangle$ is an uncorrectable error \textcolor{black}{\cite{PhysRevX.12.021049}}.

At $t=T_{g}$, the final density matrix is defined by $\hat{\rho}_{out}=|\Psi_{T_g}\rangle \langle \Psi_{T_g}|$, for each input $\hat{\rho}_{in}$, resulting in an element fidelity as
\begin{equation}
    \mathcal{F}_j= \langle {\Psi}_{j}(T_g)|\hat{\rho}_{out}|{\Psi}_{j}(T_g)\rangle
\end{equation}
which composes the realistic measurable truth table
\begin{equation}
\bar{{\rho}}=
	\begin{bmatrix}
	  {\mathcal{F}}_1 &   & &  \\
	  & {\mathcal{F}}_{2} &  & \\
	  &   & {\mathcal{F}}_{3}&\\
	  &   &  &{\mathcal{F}}_{4}
	 \end{bmatrix}
\end{equation}
The fidelity of the $C_Z$ gate is given by
\begin{equation}
\begin{aligned}
{{\mathcal{F}}}=\frac{1}{4}\text{Tr}[\sqrt{{\hat{U}}}\bar{{\rho}} \sqrt{{\hat{U}}}]^{1/2}
\end{aligned}
\end{equation}
with the standard transformation matrix for a two-qubit $C_Z$ gate
\begin{equation}
\hat{U}=
	\begin{bmatrix}
	  1 &   & &  \\
	  & -1 &  & \\
	  &   & -1&\\
	  &   &  & -1
	 \end{bmatrix}
\end{equation}

\subsection{Pulse optimization towards  active robustness}

{\it The DER pulse.-} To show a family of Rydberg quantum gates that are {\it actively} robust against any detuning error, we start by finding optimal STIRAP pulses for a two-qubit $C_Z$ gate. 
Here we optimize the gate pulses to be detuning-error robust with numerical Genetic algorithm as in our prior work \textcolor{black}{\cite{QuantumSci.Technol.8(2023)035032}}, yet replaced by using a cleverly-chosen cost function,
\begin{equation}
  J_{der} = [1-{{\mathcal{F}}}(\epsilon_\delta=0)]^2 + [{\mathcal{F}}_{\max}(\epsilon_\delta)-{\mathcal{F}}_{\min}(\epsilon_\delta)]^2.  \label{jder}
\end{equation}
Via minimizing this single-objective cost function $J_{der}$ over the range of $\epsilon_\delta\in[-\epsilon_0,\epsilon_0]$ \textcolor{black}{($\epsilon_0$ is assumed to be the maximal deviation)}, we observe that, not only the ideal gate error at $\epsilon_\delta=0$ is minimized, but also the insensitivity of gate due to the change of $\epsilon_\delta$ is significantly improved, so-called the detuning-error robust (DER) pulse. 
\textcolor{black}{It is worth emphasizing that, we expect such a DER pulse has an {\it active} robustness to any magnitude of the two-photon detuning error $\epsilon_\delta$ as long as $|\epsilon_\delta|<\epsilon_0$ is satisfied. Actually, there exists a balance between the fidelity at zero detuning (or small errors) and the robustness to the change of $\epsilon_\delta$ \cite{PhysRevLett.132.193801}. Based on the choice of $J_{der}$,
increasing the robustness should be at the expense of ideal gate fidelity \textcolor{black}{(for more discussions, see Appendix B)}.}
In contrast, the typical-optimal (TO) pulse only minimizes the ideal gate infidelity in the absence of any error by letting the cost function simply as
\begin{equation}
    J_{to} = 1-{{\mathcal{F}}}(\epsilon_\delta=0).
\end{equation}
Any deviation $\epsilon_\delta\neq 0 $ for the TO pulse can lead to a dramatic increase of the infidelity numbers [see Fig.\ref{Fig1:model}(b1-b2), the blue-dashed line], in good agreement with Fig.2 of Ref. \textcolor{black}{\cite{PRXQuantum.4.020336}}.

To our knowledge the achievable gate fidelity is fundamentally limited by the peak amplitude $\Omega_{p,c}^{\max}$ which also sets a timescale for the gate duration $T_g$. Here we fix $\Omega_{p,c}^{\max}/2\pi=150$ MHz, maintaining a high two-photon Rabi frequency $\Omega/2\pi = 5.625$ MHz, which restricts $T_g$ to be around $2.0$ $\mu$s ensured by a minimal pulse area for qubit rotation \textcolor{black}{\cite{PhysRevA.96.042306}}. Other parameters $\Delta_0/2\pi=2.0$ GHz, $B/2\pi=2.0$ GHz are kept constant throughout the paper.
That leaves a few gate parameters $(t_1,t_2,\omega_1,\omega_2)$ to be optimized. \textcolor{black}{In order to find the DER pulse, we comparably introduce two maximal deviations:
$\epsilon_0/2\pi = (0.8,0.3)$ MHz, respectively corresponding to $\epsilon_0/\Omega = (14.2\%,5.3\%)$ for showing the power of optimization algorithm under different deviation degrees.
We minimize the cost function $J_{der}$ for each $\epsilon_\delta$ within the range of $\epsilon_\delta\in [-\epsilon_0,\epsilon_0]$, which realizes an exact $C_Z$ gate with optimal pulse parameters shown in Table \ref{tab 1}.} Note that we ignore the intermediate and Rydberg decays by $\gamma_{p,r}=0$ in the optimization, and focus on the robustness of the gate pulse against detuning deviations.

The gate infidelities $1-\mathcal{F}_{der(to)}$ as a function of the detuning deviation $\epsilon_\delta$,
are displayed in Fig.\ref{Fig1:model}(b1-b2) by the black-solid (DER) and blue-dashed (TO) lines, corresponding to $\epsilon_0/2\pi=0.8$ MHz and 0.3 MHz. It is clear that the TO pulse depending on ideal optimization, benefits from a minimal infidelity $1-\mathcal{F}_{to}< 10^{-6}$ only if $\epsilon_\delta$ is very small while the infidelity numbers increase exponentially with $\epsilon_\delta$. For a large value of $|\epsilon_\delta|/2\pi = 0.8$ MHz the infidelity even reaches more than $0.02$. On the contrary, we excitingly find that the insensitivity of the robust DER pulse against the variation of $\epsilon_\delta$ has been greatly improved. See (b1), it is remarkable that the DER pulse becomes strongly robust against detuning deviations although the exact infidelity numbers stay at a relatively high level $10^{-3}\sim 10^{-2}$. 
E.g. at $|\epsilon_\delta|/2\pi=0.7$ MHz, 
the DER pulse has an infidelity of 0.00195, smaller than the TO pulse by one order of magnitude. Thus, for a wide optimization range, the DER pulse can reveal appealing advantages in improving the insensitivity of gate to detuning deviations, which may be useful for preserving the predicted gate fidelities above 0.99 for a larger error.
But, this improvement should be at the expense of low ideal gate fidelity $\mathcal{F}_{der}(\epsilon_\delta=0)\approx 0.9971$ (compared to $ \mathcal{F}_{to}\approx0.9999997$ for the TO pulse) because the cost function $J_{der}$ has incorporated the impact from two competing terms \textcolor{black}{(see Appendix B)}.

\textcolor{black}{Accounting for the fact that a real deviation $\epsilon_\delta$ is not large enough after considering various experimental imperfections, thereby we next use $\epsilon_0/2\pi = 0.3$ MHz to perform a fine numerical optimization. The results are displayed in Fig.\ref{Fig1:model}(b2). Here, in order to achieve the robust DER pulse we find the gate duration $T_g$ tends to be shortened by diminishing the wait-time between two STIRAP sequences which leads to a larger gate fidelity $\mathcal{F}_{der}(\epsilon_\delta=0)\approx 0.9998$. Moreover, as $\epsilon_\delta$ increases, the DER pulse outperforms the TO pulse over a wider range of deviations $\epsilon_\delta/2\pi> 0.114$ MHz because the time-spent in the Rydberg states has been decreased with a fine-optimization.} 
Consequently, we remark that the DER pulse with {\it active} robustness can manifest a more promising merit when the two-photon detuning deviation $\epsilon_\delta$ is large, while the TO pulse performs better only if all imperfections are very small. Using the DER pulse enables the gate implementation to be more insensitive against any magnitude of detuning errors, maintaining a stable output performance.

{\it The DER-i pulse.-} We identify that the DER pulse has promising advantages for larger two-photon detuning errors. However, maintaining this advantage in realistic experimental conditions requires the full knowledge of noise types.
As a Doppler dephasing error on the Rydberg state contributing to detuning errors, typically follows a standard Gaussian distribution \cite{PhysRevA.72.022347}, we further modify the cost function to be $J_{der-i}$ in the numerical optimization for improving its relevance on the choice of noise type,
\begin{equation}
J_{der-i} = [1-{{\mathcal{F}}}(\epsilon_\delta=0)]^2 + [{1-\bar{\mathcal{F}}(\epsilon_\delta)}]^2  \label{jder11}
\end{equation}
The first term in Eq.(\ref{jder11}) suggests maximizing the ideal gate fidelity in the absence of any noise, same as $J_{der}$. While the second term means the average gate fidelity $\bar{\mathcal{F}}(\epsilon_\delta)$ should be simultaneously maximized, where \cite{PhysRevA.107.023103} 
\begin{equation}
\bar{\mathcal{F}}(\epsilon_\delta)=\frac{\mathcal{F}(\epsilon_{\delta 1})\omega_{1}+\mathcal{F}(\epsilon_{\delta 2})\omega_{2}+...+\mathcal{F}(\epsilon_{\delta n} )\omega_{n}}{\omega_{1}+\omega_{2}+...+\omega_{n}}
\label{ave}
\end{equation}
with variables $\epsilon_{\delta i}$ uniformly extracted from $[-\epsilon_0,\epsilon_0]$.
Specifically, it is worth noting that we consider 
a set of Gaussian-type weights $\omega_{1}, \omega_{2} ,...,\omega_{n}$ with $\sum_{i=1}^{n}\omega_{i}=1$ to define the average fidelity $\bar{\mathcal{F}}(\epsilon_\delta)$. Within an arbitrary deviation range $[-\epsilon_0,\epsilon_0]$, we solve a large number of discrete fidelity numbers $\mathcal{F}(\epsilon_{\delta i})$ alongside with a Gaussian-weighted random coefficient $\omega_i$ which finally leads to a more realistic estimation for the average fidelity $\bar{\mathcal{F}}(\epsilon_\delta)$. Such setting of $J_{der-i}$ contains a tunable and noise-dependent weight $\omega_i$ that can be easily applied to other noise sources.

\textcolor{black}{The ideal infidelity $1-{\mathcal{F}}_{der-i}$ as a function of the detuning error $\epsilon_\delta$ with optimized DER-i pulses, is displayed in Fig.\ref{Fig1:model}(b1-b2) by the red-dotted lines.
We identify that the robust DER-i pulse always outperforms the DER pulse by more than one order of magnitude for small detuning errors, and even satisfies $\mathcal{F}_{der-i}>0.9999$ (ideal fidelity) due to the fine optimization. This is expected because the Gaussian sampling of $\epsilon_\delta$ in defining $\bar{\mathcal{F}}(\epsilon_\delta)$ intrinsically makes the real detuning error felt by atoms smaller, hence the gate fidelity is relatively large when $\epsilon_\delta$ is small. The DER-i pulse tends to achieve a larger value of fidelity for small detuning errors, consequently the robustness of which suffers from a small decrease as compared with the DER pulse. But now the DER-i pulse becomes the best choice for reasonable detuning errors by taking account of both a large gate fidelity and the robustness to detuning imperfections. We find it can achieve an infidelity of $1-\mathcal{F}_{der-i}< 10^{-4}$ even for $|\epsilon_\delta|/2\pi$ up to 0.1 MHz when the optimization range is fine. As for a wide optimization case, the robustness of DER-i pulse can also be improved by restricting the fluctuation of infidelity numbers within $10^{-4}\sim 10^{-2}$ for any $\epsilon_\delta$, 
significantly better than the TO pulse. }

\section{Robust pulses under different noise sources}

\begin{figure}
    \centering
    \includegraphics[scale=0.37]{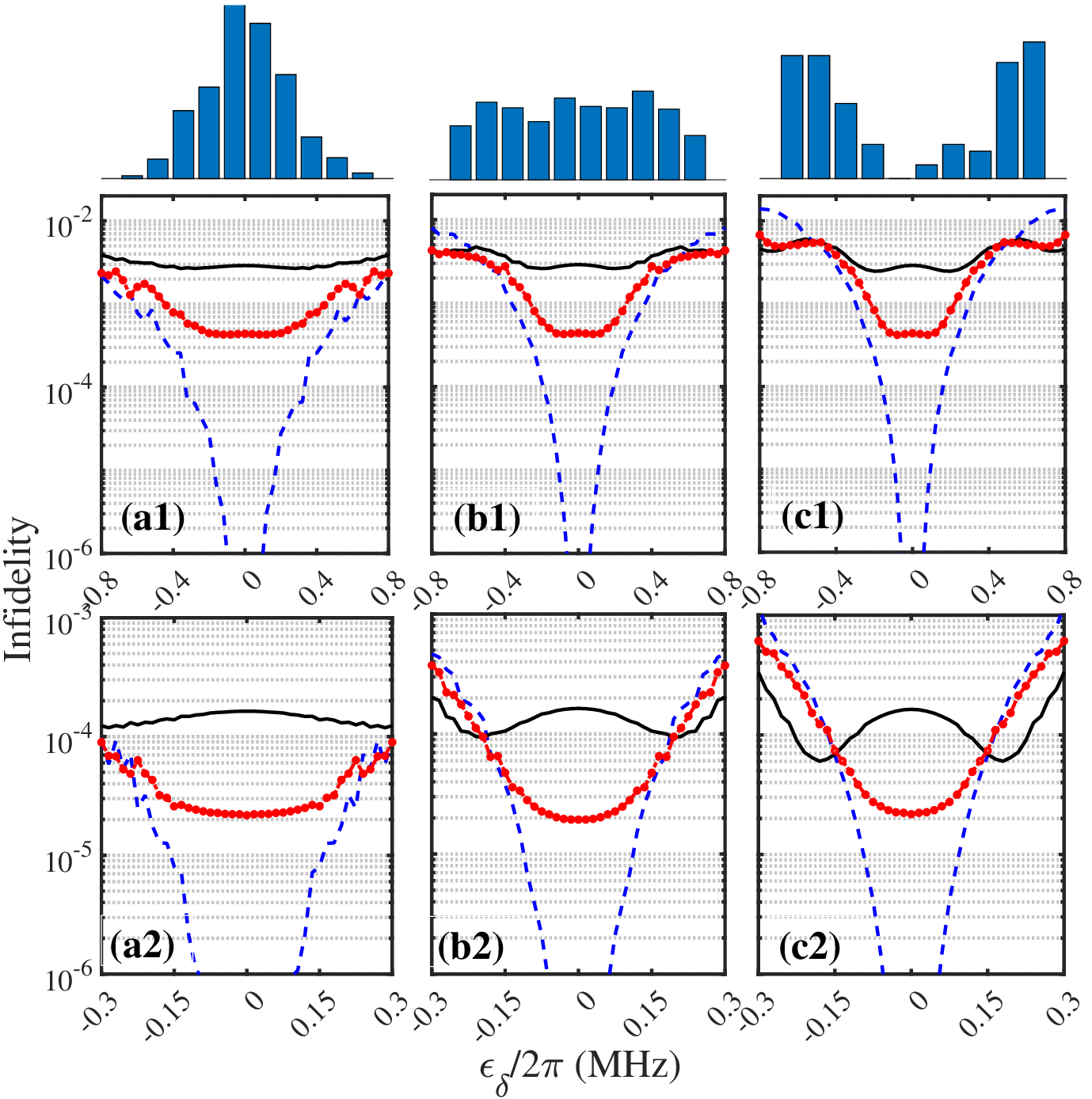}
    \caption{The infidelity of the DER(black-solid), the DER-i(red-dotted) and the TO(blue-dashed) pulses under different types of noise sources. From left to right, for each $\epsilon_\delta$, the real sampling numbers are randomly distributed within the range of $[-\epsilon_\delta,\epsilon_\delta]$, satisfying Gaussian, Uniform and U-shaped distributions with respective profiles schematically shown on the top. Cases of $\epsilon_0/2\pi = 0.8$ MHz and 0.3 MHz are given in (a1-c1) and (a2-c2).
    Each point presents an average over 500 random samplings. No spontaneous decay is considered in the calculation.}
    \label{Fig2:threedistribution}
\end{figure}

Accounting for the fact that the realistic detuning error caused by versatile experimental imperfections is essentially unknown, without loss of generality, we choose three different representations of the error distributions: Gaussian, Uniform, U-shaped, individually for low, middle and high noise sources. The infidelity is shown as a function of detuning error $\epsilon_\delta$ in Fig.\ref{Fig2:threedistribution} with same set of optimal pulse parameters (see Table I, rows 3-7). For a certain value $\epsilon_\delta$, the realistic error number in single sampling is randomly obtained from a specific error function (top) and we display the average result after significant samplings. 

Seen from Fig.\ref{Fig2:threedistribution}(a1-c1), for a wide error range $\epsilon_\delta/2\pi\in[-0.8,0.8]$ MHz, the DER pulse always presents a steady insensitivity to the variation of $\epsilon_\delta$ no matter what the noise distribution is; while the TO pulse does not show any robustness and is extremely sensitive to $\epsilon_\delta$. This is expected because the DER pulse has exhibited the ability to overcome the influence from detuning deviation resulting in an {\it active} robustness yet at the expense of gate fidelities. Moreover, note that the DER-i pulse involving the consideration of weighted average fidelity has a much improved insensitivity as compared to the TO pulse, while also benefiting from a higher fidelity ($\mathcal{F}_{der-i}(\epsilon_\delta=0) \approx 0.9996 $) than the DER pulse.
As accounting for the noise types, we verify that, when the noise is relatively low forming as a Gaussian distribution, the DER pulse has no advantages as compared to the TO pulse because it sustains a poor infidelity above $10^{-3}$ as $\epsilon_\delta$ varies. The DER pulse becomes a better choice only if the detuning noise is significantly large. E.g. by applying the Uniform or the U-shaped noises, we find the DER pulse is best for $|\epsilon_\delta|/2\pi>0.6$ MHz. In addition, we expectedly find the improved DER-i pulse outperforms the DER pulse by bringing a more acceptable gate fidelity when $|\epsilon_\delta|/2\pi<0.4$ MHz, which achieves {$1-\mathcal{F}_{der-i}\approx 4\times10^{-4}$}.
Beyond that, although it behaviors similarly to the TO pulse, yet the DER-i pulse can show much better robustness to the detuning deviations.
Therefore we remark that, under a relatively strong noise environment \cite{PhysRevA.97.053803}, the DER-pulse protocols can outperform the TO protocol by exhibiting an exactly steady output with perfect robustness. An acceptable gate fidelity value above $ 0.99$ remains attainable for a large detuning error. Moreover, the DER-i pulse is also significantly better than the TO pulse by providing both a high fidelity and good robustness.

\textcolor{black}{Similar results under fine optimization are summarized in Fig.\ref{Fig2:threedistribution}(a2-c2). For a Gaussian-type noise the DER pulse has no advantages in fidelity because the average error felt by atoms is reduced due to the Gaussian distribution. And the DER-i pulse only outperforms the TO pulse at very large values of $\epsilon_\delta$. While for Uniform or U-shaped noises, we observe that, the DER pulse always preserves the most stable output when $\epsilon_\delta$ is varied, and the achievable infidelity has been lowered to be around $1-\mathcal{F}_{der}\approx 10^{-4}$ over the complete optimization range. In contrast, the DER-i pulse is more qualitatively similar to the TO pulse providing a larger gate fidelity. E.g. its infidelity number is close to $\sim10^{-5}$ for small deviations. However, its {\it active} robustness to the error fluctuations is slightly weak because the infidelity  exponentially increases as $\epsilon_\delta/2\pi \gtrsim 0.12$ MHz.}

To conclude we have presented two different robust pulses DER and DER-i which provide advanced improvements in achieving higher gate fidelities and stronger robustness, especially in a noisy environment. By comparing the pulse performance we show the fine optimization with small error deviations is more favored by us due to the existence of a lower average fidelity.

\section{Potential Applications to realistic noises}

In a realistic experiment, two common major error sources that severely limit the fidelity of quantum gate operations, are the finite atom temperature \textcolor{black}{\cite{PhysRevLett.123.230501}} and the laser amplitude noise \textcolor{black}{\cite{PhysRevA.102.042607}}. The former causes significant motional dephasing of the ground-Rydberg transitions. And the latter not only leads to the fluctuated Rabi frequencies, but will also affect the two-photon detuning by inducing ac Stark shifts \textcolor{black}{\cite{PhysRevA.95.043429}}. Especially for gate schemes relying on the optimal control of pulse shapes this laser-amplitude derivation(the latter) 
can significantly affect the observed gate fidelity \textcolor{black}{\cite{PhysRevA.105.042404}}. Here we develop the appealing gates with {\it active} robustness to the errors associated with the variation in two-photon detuning. That means the gates are in principle more robust against any type of errors affecting the two-photon detuning.

In this section, we evaluate the realistic gate performance under two major error sources.
Other technical imperfections that passively influence the gate, will be analyzed in \textcolor{black}{Appendix C}.

\subsection{Doppler dephasing error}

So far the atomic thermal motion-induced Doppler dephasing is the most dominant error source in experiment, which ultimately limits the gate performance
\textcolor{black}{\cite{PhysRevLett.121.123603}}. Atoms, whether they are warm or cold, are not stationary in the trap. The random velocity of trapped atoms would lead to the dephasing of ground-Rydberg excitation, manifested as a modification to the phase of laser Rabi frequencies $\Omega_p$ and $\Omega_c$, denoted by \cite{PhysRevApplied.13.024008}
\begin{equation}
    \Omega_{p}(t)\to \Omega_{p}(t) e^{i\epsilon_{\delta_p}t}, \Omega_{c}(t)\to \Omega_{c}(t) e^{i\epsilon_{\delta_c}t} \label{dopper}
\end{equation}
where $\epsilon_{\delta_{p(c)}} = \vec{k}_{p(c)}\cdot\vec{v}$ is the Doppler shift with $\vec{k}_{p,c}$ two participating laser wavevectors and $\vec{v}$ the atomic random velocity extracted from a Gaussian distribution with standard deviation $v_{rms}=\sqrt{k_BT/m}$ \textcolor{black}{\cite{PhysRevA.72.022347}}, where $T$ is the temperature and $m$ is the mass of atoms. Here we choose the laser wavelengths for the two-photon transition of $\lambda_p = 420$ nm and $\lambda_c = 1013$ nm [see Fig.\ref{Fig1:model}(a)]. In order to diminish the Doppler dephasing, we use counterpropagating lasers instead of orthogonal ones which gives a minimal two-photon wavevector $k_{eff} =2\pi(1/\lambda_p-1/\lambda_c) = 8.76\times 10^6 $ m$^{-1}$. For a finite temperature $T$, this modification of laser phase equivalently means the realistic detuning fluctuation of excitation lasers felt by the atoms is a random variable which are $\epsilon_\Delta = \epsilon_{\delta_p}= k_pv$ and $\epsilon_\delta =\epsilon_{\delta_p}-\epsilon_{\delta_c}= k_{eff}v$.

\begin{figure}
    \centering
    \includegraphics[scale=0.36]{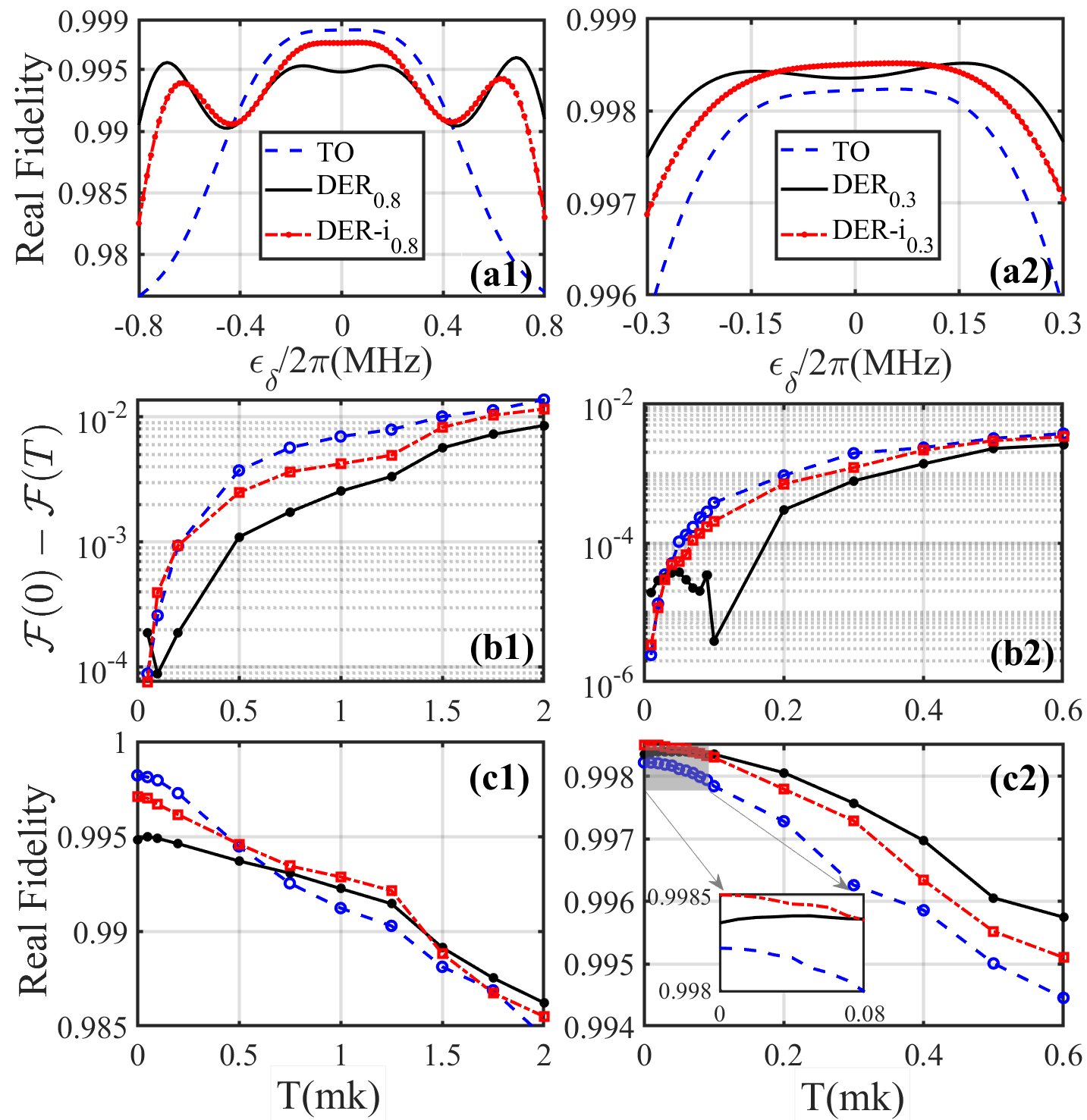}
    \caption{The realistic gate performance with the optimal TO(dashed-blue), DER(black-solid), and DER-i(dash-dotted red) pulses, corresponding to different optimization ranges $\epsilon_0/2\pi=0.8$ MHz(left column) and $0.3$ MHz(right column). (a1-a2) Real fidelity versus the detuning error $\epsilon_\delta$. (b1-b2) Infidelity as a function of the atomic temperature $T$. (c1-c2) The realistic gate fidelity versus $T$. Inset of (c2): The small range of $T\in[0,80]$ $\mu$K is amplified. 
    Here, the spontaneous decays with rates $\gamma_{p} = 8.4746$ MHz, $\gamma_r=2.7$ kHz from the intermediate state $|p\rangle$ and the Rydberg state $|r\rangle$, are considered in the calculation, and other imperfections are not included. All optimized gate parameters are given by Table \ref{tableI}. In (b-c), each point represents an average over 500 random samplings. 
    }
    \label{Fig3:doppler}
\end{figure}

\textcolor{black}{For demonstrating the improved robustness against Doppler dephasing errors, we first display the real gate fidelity for different detuning errors $\epsilon_\delta$, analogously to Fig.\ref{Fig1:model}(b1-b2) but 
in the presence of spontaneous decays. By varying $\epsilon_\delta$, if $\epsilon_0/2\pi = 0.8$ MHz (wide optimization), it is clear that the DER and DER-i pulses can counterintuitively achieve a high fidelity $\sim 0.995$ for larger detuning deviations; however, which is at the cost of a low average gate fidelity under small deviations. Because there exists a balance between them and an improved robustness requires the cost of ideal gate fidelities. On the contrary, as for a sufficiently fine optimization, we see that both DER and DER-i pulses have dominant advantages wholly outperforming the TO pulse for any $\epsilon_\delta$. Because the proposed gates utilizing optimized DER or DER-i pulses spend less time in the intermediate excited states and are thus suffering from a smaller decay error than the TO pulse. In addition, by comparing DER and DER-i it is expected that the DER-i pulse shows the best fidelity $F_{der-i}\approx 0.9985$ for $|\epsilon_\delta|/2\pi<0.12$ MHz even outperforming the DER pulse there, because the optimization for DER-i pulse inherently has included a consideration of Gaussian-weighted noise distribution, leading to a smaller effect on the average fidelity at small $\epsilon_\delta$ values.}

The real gate infidelity caused by residual thermal motion of atoms is revealed by calculating the relationship between $\mathcal{F}(0)-\mathcal{F}(T)$ and the atomic temperature $T$, as plotted in Fig.\ref{Fig3:doppler}(b1-b2). To quantify the Doppler effect, we compare the results by using the TO, DER, and DER-i pulses for different optimization ranges. Accordingly, we choose a reasonable range for the atomic temperature variation to be $T\in [0,2.0]$ mK ($k_{eff}v_{rms}/2\pi \lesssim 0.6097  $ MHz) and $T\in[0,0.6]$ mK ($k_{eff}v_{rms}/2\pi  \lesssim 0.3339 $ MHz)
by considering the long-tail feature of Gaussian function. As $T$ increases, we observe that both the DER and DER-i pulses benefit from a smaller dephasing error for any $T$, while the DER pulse is even better because of the significantly-improved robustness against Doppler errors.
Furthermore, to show the gate performance in a more realistic condition, we calculate the gate fidelity by involving the effects of spontaneous decays as well as a finite temperature, as display in
Fig.\ref{Fig3:doppler}(c1) and (c2). As for a wide optimization, the DER and DER-i pulses have no advantages at low temperatures, because the TO pulse is expected to perform best when the imperfection is sufficiently small.
However, as $T$ increases {\it e.g.} $T>0.5$ mK, we remark that the gate fidelity using the TO pulse is dramatically decreased, while the DER and DER-i pulses can provide higher fidelities there which strongly confirms the improved robustness against Doppler errors with them.

\textcolor{black}{Finally, we calculate the real gate fidelity in the presence of Doppler errors by using fine-optimization pulses. Fig.\ref{Fig3:doppler}(c2) shows the gate fidelity of three pulses over a relatively small range of imperfections {\it i.e.} $T\in [0,0.6$] mK. In general
we find the DER pulse can provide a noticeable but modest fidelity improvement with increasing temperature, while the TO pulse always performs worst. E.g. at $T = 0.2$ mK the real fidelity of three pulses is $(\mathcal{F}_{der},\mathcal{F}_{der-i},\mathcal{F}_{to} )= (0.9981,0.9978,0.9972)$. As taking into account of very small temperatures $T<80$ $\mu$K we see the DER-i pulse turns to perform best, {\it e.g.} $(\mathcal{F}_{der},\mathcal{F}_{der-i},\mathcal{F}_{to} )= (0.9983,0.9985,0.9982)$ for $T=20 $ $\mu$K. Because, the optimization for DER-i pulse has incorporated with the Gaussian-noise distribution in which the small detuning error plays the most vital role.}
Therefore, the optimized DER and DER-i pulses can provide a better performance than the TO pulse as long as $T$ is moderately varied within the optimization range. These results are able to relax the temperature constraint for trapped atoms in optical tweezers and greatly protect the coherence of quantum gate operation from the finite temperature effect \cite{PhysRevApplied.18.044042}.

\subsection{Ac Stark shift error}

\begin{figure}
    \centering
    \includegraphics[scale=0.42]{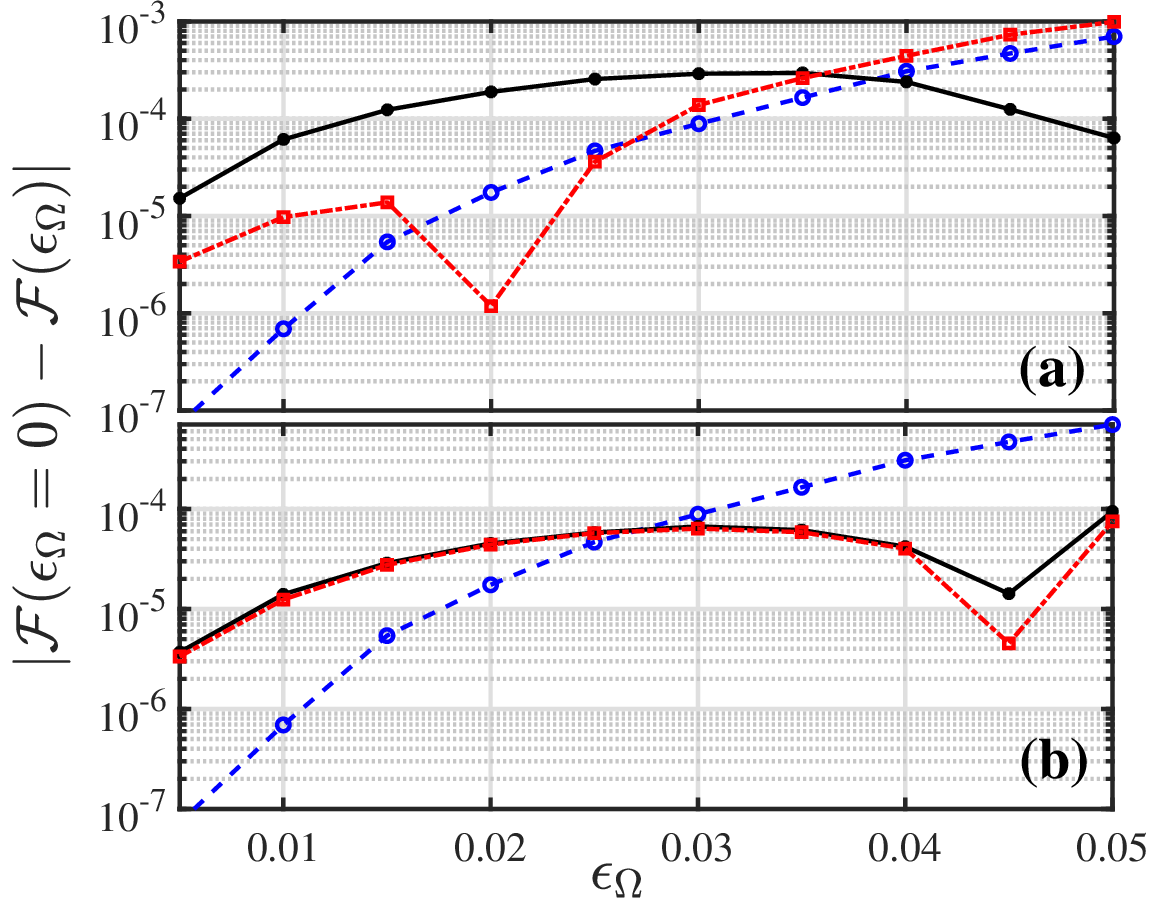}
    \caption{(a-b) Estimation of the ac Stark shift error induced by the laser amplitude deviation $\epsilon_\Omega$ at 
    $T=0$, corresponding to different  optimization ranges $\epsilon_0/2\pi = 0.8$ MHz and 0.3 MHz. Here the DER-i pulse is re-optimized with a set of Uniform-type weights (see Eq.\ref{ave}), resulting in the optimized parameters given in Table I (denoted by dark green). For each $\epsilon_{\Omega}$, the Stark shift error $\epsilon_\delta$ is randomly extracted from $\max[\frac{\Omega_p^2+\Omega_c^2}{2\Delta_0}]\times [-\epsilon_\Omega,+\epsilon_\Omega]$ adding to the two-photon detuning, which affects the gate fidelity. The average result after 500 samplings is presented. The linetype is same as in Fig.\ref{Fig3:doppler}.}
    \label{Fig4:laseramplitude}
\end{figure}

In the following, we proceed by showing that the DER and DER-i pulses also have some robustness against the ac Stark shift caused by the laser amplitude deviation. Amplitude fluctuations of the excitation lasers will affect the Rabi frequencies $\Omega_p(t)$ and $\Omega_c(t)$ inducing a typical laser amplitude noise (see Appendix C). More crucially, it also correlates an extra detuning error $\epsilon_\delta$.
Because, the laser amplitude fluctuations \textcolor{black}{\cite{PhysRevResearch.5.033052}} can result in an uncertainty in Rabi frequencies as
\begin{equation}
 \Omega_{p}(t)\rightarrow\Omega_{p}(t)(1+ \epsilon_{\Omega_{p}}),\Omega_{c}(t)\rightarrow\Omega_{c}(t)(1+ \epsilon_{\Omega_{c}})
 \label{ome}
\end{equation}
For a two-photon transition system with a large intermediate detuning $|\Delta_0|\gg\Omega_{p,c}$, a significant ac Stark shift will occur to the two-photon resonance $\delta_{ac} = \delta_0+\frac{\Omega_c^2-\Omega_p^2}{4\Delta_0}$(remember $\delta_0=0$), which evolves with time since $\Omega_{p}(t),\Omega_{c}(t)$ are both time-dependent \textcolor{black}{\cite{PhysRevB.91.224506}}. While replacing by the fluctuated Rabi frequencies, the realistic ac shift $\delta_{ac}$ takes a more complex form as
\begin{equation}
    \delta_{ac} = \delta_{ac,0}  + \frac{\Omega_{p}^{2}}{2\Delta_0}\epsilon_{\Omega_{p}}-\frac{\Omega_{c}^{2}}{2\Delta_0}\epsilon_{\Omega_{c}}+\mathcal{O}(\epsilon_{\Omega_{p}}^2,\epsilon_{\Omega_{c}}^2)
\end{equation}
where the known (first) term $\delta_{ac,0}$ can be sufficiently compensated by a fixed detuning in experiment \textcolor{black}{\cite{PhysRevLett.129.200501}}. Nevertheless, the unknown (second and third) terms would induce an unknown two-photon detuning error, which is described by
\begin{equation}
  \epsilon_{\delta} =  \frac{\Omega_{p}^{2}}{2\Delta_0}\epsilon_{\Omega_{p}}-\frac{\Omega_{c}^{2}}{2\Delta_0}\epsilon_{\Omega_{c}}   
\end{equation}
where we have ignored the higher-order terms. Here, the maximal detuning deviation $\epsilon_{\delta,\max}$ induced by ac Stark shifts can be roughly calculated by 
\begin{equation}
    \epsilon_{\delta,\max} = \max[\frac{\Omega_p^2+\Omega_c^2}{2\Delta_0}]\times \epsilon_\Omega
    \label{max}
\end{equation}
at $\epsilon_\Omega = \epsilon_{\Omega_p} = -\epsilon_{\Omega_c}$, 
 which is linear with the increase of $\epsilon_\Omega$. For a large amplitude deviation $\epsilon_\Omega = 0.05$, the maximal detuning error is about $\epsilon_\delta \approx 2\pi\times 0.32$ MHz.

To quantify the robustness of different optimal pulses to this extra detuning deviation $\epsilon_\delta$ ({\it i.e.} the ac Stark shift error), relevant results are summarized in Fig.\ref{Fig4:laseramplitude}. 
We calculate the absolute infidelity 
$|\mathcal{F}(\epsilon_\Omega=0)-\mathcal{F}(\epsilon_\Omega)|$(without the decay error) 
for different values of $\epsilon_\Omega$ at $T=0$. For each $\epsilon_\Omega$ the induced Stark error is randomly obtained from $\epsilon_\delta\in [-\epsilon_{\delta,\max},+\epsilon_{\delta,\max}]$ according to Eq.(\ref{max}), serving as a two-photon detuning error for state $|r\rangle$. We treat $\epsilon_\Omega$ only 
as a quantity to measure the strength of Stark error while keeping $\Omega_{p}(t)$ and $\Omega_c(t)$ unvaried for avoiding the amplitude error.
Since a typical amplitude deviation $\epsilon_\Omega$, not only induces a laser amplitude error but also correlates an extra two-detuning error $\epsilon_\delta$ via ac Stark shift effect. The former will be discussed in Appendix C.

\textcolor{black}{Here in Fig.\ref{Fig4:laseramplitude}(a-b), we observe that, the TO pulse does not show any robustness to the ac Stark shift error and increases exponentially with $\epsilon_\Omega$ as expected. While for two robust pulses DER and DER-i, the improvement of robustness is explicit. 
In the case of wide optimization, see Fig.\ref{Fig4:laseramplitude}(a), we find the DER pulse is more robust to the Stark error especially for large deviations because of its inherent robustness to large detuning errors. Here the DER-i pulse is qualitatively similar to the TO pulse without having noticeable advantages.
As turning to the fine-optimization case, we admit that two robust pulses are relatively worse than the TO pulse in the range of small detuning deviations because the improvement of robustness has to loss some gate infidelities. However, both of them reveal sufficient robustness against the ac Stark shift error for $\epsilon_\Omega > 0.025$, enabling the gate infidelity to be staying below $\sim 10^{-4}$, which outperforms the TO pulse by orders of magnitude.
This is achieved by the fine optimization algorithm that allows the robust pulses to have both a large gate fidelity and strong robustness.}

In Sec. V.A and V.B, we have shown the {\it active} robustness of optimized DER and DER-i pulses against two realistic detuning errors arising from Doppler dephasing and ac Stark shift. As for the Doppler shift which serves as a significant two-photon detuning error, the robustly optimized pulses DER and DER-i can always outperform the non-robust TO pulse by using a fine-optimization method. While for the ac Stark shift error induced by the laser amplitude deviation, a Stark shift-robust version only exists for the larger amplitude deviations.
Because two robust pulses are inherently obtained at the expense of gate fidelities at small detuning errors.
While, in general, the DER and DER-i pulses can show stronger robustness to the change of amplitude deviations, {\it e.g.} for $\epsilon_\Omega<0.05$, the ac Stark shift error of both robust pulses can stay within $10^{-5}\sim 10^{-4}$, significantly better than the TO pulse which exponentially increases from $10^{-7}$ to $10^{-3}$.



\section{Discussion and Summary}

We present a scheme of high-fidelity adiabatic $C_Z$ gates natively having {\it active} robustness to one type of errors, {\it i.e.}, the Rydberg two-photon detuning error. We mainly focus on optimizing the laser pulses to be insensitive towards the deviation of two-photon detuning, which may originate from versatile experimental imperfections, such as atomic temperature effect or laser amplitude deviation. Based on the use of a cleverly-modified cost function that targets at minimizing the fluctuation of infidelity(average infidelity) in a certain error range, we observe that, the gate error due to Doppler shifts and laser amplitude deviations can be reduced, although at the expense of a slightly lower ideal fidelity at small detunings. Additionally, the optimized laser pulse can be more effective by accounting for the type of error distributions. One example is when the DER-i pulse contains Gaussian-weighted coefficients in defining the average fidelity function, it can outperform other two pulses (DER and TO) with an explicitly higher gate fidelity for lower atomic temperatures. In addition, the newly-proposed DER and DER-i pulses under a fine optimization have higher gate fidelities for any temperature, which significantly relaxes the atomic temperature constraint in a Rydberg experiment by reducing the dephasing effect from atomic velocity and position fluctuations \textcolor{black}{\cite{PhysRevLett.132.043201}}.

While we still admit that, the robust  pulses (DER and DER-i) do not explicitly have enhanced and {\it active} robustness to other types of noises, such as the intermediate detuning error, inhomogeneity in Rabi frequencies, laser amplitude and phase noises, and the fluctuated interatomic interactions (see Appendix C). Especially the laser phase noise manifests as a time-dependent parameter fluctuation which contributes to the gate error at a comparable level as the laser amplitude noise \textcolor{black}{\cite{PhysRevA.107.042611,PhysRevA.97.053803}}. Finally we believe that, our results provide a more prospective method for achieving high-fidelity Rydberg gates, combining with the promising merits of {\it active} robustness to any imperfection associated with the two-photon detuning deviations. An actively noise-robust pulse will inspire further experimental improvements focusing on the suppression of one type of errors for some important tasks in the field of large-scale quantum computation \textcolor{black}{\cite{Naturecommunications(2022)13.4657,PhysRevApplied.19.064038,PhysRevA.109.042615,PhysRevApplied.20.034019}}.

\begin{acknowledgments}
We acknowledge financial support from the National Natural Science Foundation of China under Grants No. 12174106, No. 11474094 and  No. 11104076, by the Science and Technology Commission of Shanghai Municipality under Grant No.18ZR1412800.
\end{acknowledgments}

\appendix

\section{Effective dynamics for $C_Z$ gates}

\begin{figure*}
    \centering
\includegraphics[scale=0.68]{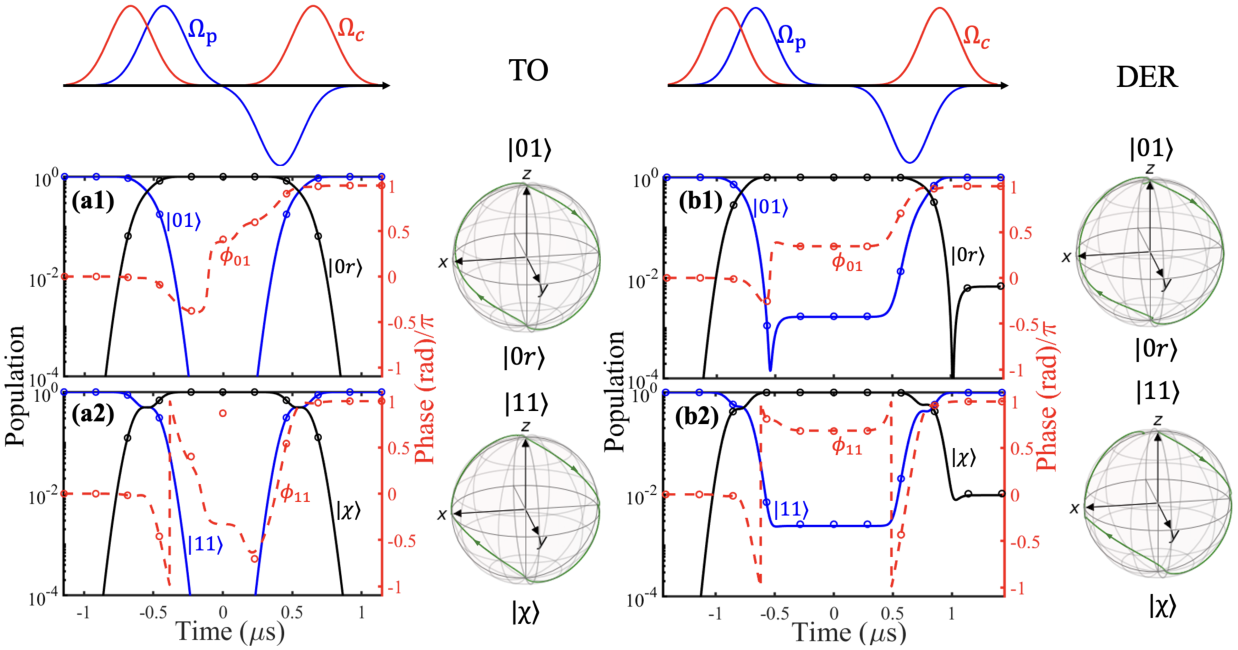}
    \caption{The population dynamics associated with the effective two-level model and the corresponding phase evolution of initial states (a1) $|01\rangle$ and (a2) $|11\rangle$ with the TO pulse, where no decay is considered. Similarly, (b1-b2) present the case with the DER pulse for $\epsilon_0/2\pi=0.8$ MHz. Other intermediate-state population is not shown. For the TO pulse, 
    the population returns back with the accumulated phases $\phi_{10} \approx 1.00000\pi$ and $\phi_{11} \approx 0.99999\pi$, resulting in $2\phi_{01} - \phi_{11} \approx 1.00001\pi$. Analogously, for the DER pulse, $\phi_{10} \approx 1.00000\pi$ and $\phi_{11} \approx 0.99902\pi$ arising 
     $2\phi_{01} - \phi_{11} \approx 1.00098\pi$. Good agreement is observed between the effective two-level systems(circles) and the originally full dynamics(lines).}
    \label{Fig5:effective}
\end{figure*}

In this section, we demonstrate the physics associated with the implementation of a $C_Z$ gate. We start with the single-atom frame described by a three-level non-fluctuated Hamiltonian in the $\{|1\rangle,|p\rangle,|r\rangle\}$ basis
\begin{equation}
{{\hat{H}}}=\frac{\Omega_p}{2}|p\rangle \langle 1|+\frac{\Omega_c}{2}|r\rangle \langle p|+\text{H.c.}- \delta_0|r\rangle \langle r|-\Delta_0|p\rangle \langle p|  
\end{equation}
in which we use symbol $\delta_0$ for a small two-photon detuning (note that $\delta_0=0$ in the maintext). Due to the large detuning condition $|\Delta_0|\gg \Omega_{p,c},\delta_0$, we can perform adiabatic elimination for state $|p\rangle$ \cite{PhysRevA.89.032334} which leads to the effective two-level dynamics, as
\begin{equation}
{{\hat{H}}}_{eff}=\frac{\tilde{\Omega}}{2}|r\rangle \langle 1| +\text{H.c.} + \frac{\Omega_{p}^2}{4\Delta_0}|1\rangle \langle 1| + (\delta_0 +\frac{ \Omega_{c}^2}{4\Delta_0})|r\rangle \langle r|
\label{eff}
\end{equation}
where $\tilde{\Omega}=\frac{\Omega_{p}\Omega_{c}}{2\Delta_0}$ is the effective Rabi frequency.

When the input state is $|01\rangle$(equally for $|10\rangle$) with only one excited atom, 
the dynamics is given by the coupling of single atom, rewritten as
\begin{equation}
{{\hat{H}}}_{01}=\frac{\tilde{\Omega}}{2}|0r\rangle \langle 01| +\text{H.c.} + \delta^\prime|0r\rangle \langle 0r|
\label{H_eff_01}
\end{equation}
with the two-photon detuning $\delta^\prime = \delta_0+\delta_{ac}=\delta_0 + \frac{\Omega_{c}^2-\Omega_p^2}{4\Delta_0}$, forming a two-level system with Rabi frequency $\tilde{\Omega}$ and detuning $\delta^\prime$. 

Whereas, it becomes a little more complicated for the $|11\rangle$ state.
There are six symmetric states $\{|11\rangle,\frac{|1p\rangle+|p1\rangle}{\sqrt{2}},|pp\rangle,\frac{|1r\rangle+|r1\rangle}{\sqrt{2}},\frac{|pr\rangle+|rp\rangle}{\sqrt{2}},|rr\rangle\}$ related to this problem. After adiabatic elimination of $|p\rangle$ the initial Hamiltonian for state $|11\rangle$ can be described by
\begin{equation}
{{\hat{H}}}_{11}= {{\hat{H}}}_{eff}\otimes \hat{I} + \hat{I}\otimes{{\hat{H}}}_{eff} + B|rr\rangle\langle
rr|
\end{equation} 
which is equivalent to the matrix form in the $\{|11\rangle,|\chi\rangle ,|rr\rangle\}$ basis with $|\chi\rangle = \frac{|1r\rangle+|r1\rangle}{\sqrt{2}}$,
\begin{equation}
{{\hat{H}}}_{11} = 
	\begin{bmatrix}
	\frac{\Omega_{p}^2}{2\Delta_0}& \frac{\sqrt{2}\tilde{\Omega}}{2}& 0 \\
	 \frac{\sqrt{2}\tilde{\Omega}}{2} & \delta_{0}+ \frac{\Omega_{c}^2}{2\Delta_0}&\frac{\sqrt{2}\tilde{\Omega}}{2} \\
	  0 & \frac{\sqrt{2}\tilde{\Omega}}{2} & 2(\delta_{0}+\frac{\Omega_{c}^2}{4\Delta_0})+B
	 \end{bmatrix}
\end{equation}
After eliminating $|rr\rangle$ within the strong Rydberg blockade regime, we arrive at an effective coupling between 
$|11\rangle$ and $|\chi\rangle$ as a similar two-level system
\begin{equation}
{{\hat{H}}}_{11} = \frac{\sqrt{2}\tilde{\Omega}}{2}|\chi\rangle\langle11| +\text{H.c.} +\delta^{\prime\prime}|\chi\rangle\langle\chi|
\end{equation}
with enhanced Rabi frequency $\sqrt{2}\tilde{\Omega}$ and the detuning
$\delta^{\prime\prime} = \delta^{\prime}-\frac{\tilde{\Omega}^2}{\delta_{0}+\frac{\Omega_{c}^2}{\Delta_0}+2B} $.

\begin{figure*}
    \centering
    \includegraphics[scale=0.85]{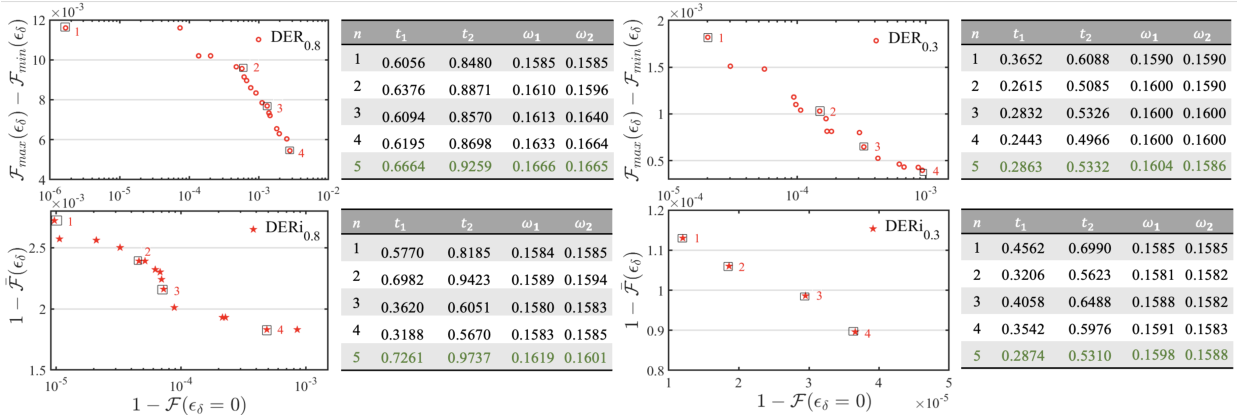}
    \caption{Results from multi-objective optimization for the robust DER and DER-i pulses, under different detuning deviations $\epsilon_0/2\pi = 0.8$ MHz (right panels) and 0.3 MHz (left panels). 
    Sufficient Pareto points are found standing for a good trade-off between two objective functions. Tables on the right illustrate four representative pulse parameters $(t_1,t_2,\omega_1,\omega_2)$(in units of $\mu$s), corresponding to the circled points in the figure. The last row (highlighted in dark green) presents the optimal parameters based on single-objective function, same as in Table I.}
    \label{Fig6:twoobject}
\end{figure*}

To verify the effective two-level dynamics, we now numerically plot the implementation of a two-qubit $C_Z$ gate by taking the specific TO and DER pulses as examples. In Fig.\ref{Fig5:effective}(a1-a2), the dynamical evolutions of initial $|01\rangle$ and $|11\rangle$ states, the intermediate $|0r\rangle$ and $|\chi\rangle$ states, 
combined with their accumulated phases $\phi_{01},\phi_{11}$, are presented for the TO pulse. Remarkably, via the use of double STIRAP pulse sequences(see top), both $|01\rangle$ and $|11\rangle$ return to their initial positions on the Bloch sphere forming a completely enclosed area \cite{PhysRevA.69.050301}. This implies that the sufficient population exchange between $|01\rangle$(or $|11\rangle$) and $|0r\rangle$(or $|\chi\rangle$) appears by obeying the full adiabatic transfer and other intermediate-state population has been strongly suppressed. The resulting infidelity achieves $1-\mathcal{F}_{to} \approx  3\times 10^{-7}$ at zero detuning (see Table I). Here, note that $|00\rangle$ is uncoupled and therefore, accumulates no phase. State $|01\rangle$ and $|10\rangle$ are equivalent by symmetry so $\phi_{01}=\phi_{10}$. Hence, we clearly see the population returning back to the desired state $|01\rangle$ and $|11\rangle$ after performing the double STIRAP pulse sequences in which the total phase accumulation is $\phi_{00}-\phi_{01}-\phi_{10}+\phi_{11}\approx -1.00001\pi$ which realizes the $C_Z$ gate. 

\textcolor{black}{For comparison, we also study the population dynamics for the DER pulse in Fig. \ref{Fig5:effective}(b1-b2). We remark that the DER pulse requires a longer time-spent on intermediate Rydberg states $|0r\rangle$ and $|\chi\rangle$, giving rise to a less sufficiently adiabatic transfer due to the staying of Rydberg population. In this case the DER pulse becomes inherently more robust against the detuning errors on Rydberg states, and achieves $1-\mathcal{F}_{der}\approx 2.9\times 10^{-3}$ when the detuning deviation is zero. Based on the Bloch sphere version we also clearly see the adiabatic transfer is incomplete for the $|01\rangle$ and $|11\rangle$ states, and hence confirm that the TO pulse can outperform the DER pulse by having a smaller infidelity at zero detuning. }


\section{Balanced optimization for DER and DER-i pulses}

We now turn to study the importance of a specifically-designed cost function that can balance the competing effects from ideal fidelity and gate robustness. To find such a robust pulse, we minimize the single cost function using the Genetic optimization algorithm. See Eq.(\ref{jder}) and (\ref{jder11}), 
the minimization of the first term $1-\mathcal{F}(\epsilon_\delta=0)$ is obtained when the ideal fidelity in the absence of any decay, is maximized. While the second terms $\mathcal{F}_{\max}(\epsilon_\delta)-\mathcal{F}_{\min}(\epsilon_\delta)$ or $1-\bar{\mathcal{F}}(\epsilon_\delta)$ are contributed by minimizing the fluctuation of exact fidelity numbers in a certain error range $[-\epsilon_0,\epsilon_0]$. In the maintext, based on a set of optimal pulse parameters $(t_1,t_2,\omega_1,\omega_2)$, a single cost function $J_{der}$ or $J_{der-i}$ has been minimized by comparing the competing effects from two terms, so-called the single-objective optimization.

In order to show the competing effect of two terms which arises a clear balance between the realistic gate fidelity and the robustness to detuning deviations, here we discuss the results making use of multi-objective optimization \textcolor{black}{\cite{PsychologyPress(2014)}}, where the cost functions are replaced by
\begin{eqnarray}
    J_{der1} &=& 1-\mathcal{F}(\epsilon_\delta=0) \nonumber\\
    J_{der2} &=& \mathcal{F}_{\max}(\epsilon_\delta) - \mathcal{F}_{\min}(\epsilon_\delta)
\end{eqnarray}
for the DER pulse, and by
\begin{eqnarray}
    J_{der1-i} &=& 1-\mathcal{F}(\epsilon_\delta=0) \nonumber\\
    J_{der2-i} &=& 1-\bar{\mathcal{F}}(\epsilon_\delta)
\end{eqnarray}
for the DER-i pulse. By utilizing the multi-objective function in the numerical Genetic algorithm, we
observe that, there is a very obvious competition between the two objective functions. This finding further confirms that if we want to improve the pulse robustness, the ideal fidelity at small deviations needs to be sacrificed.

In the multi-objective optimization, we introduce the Pareto solution to find the most adaptive pulse parameters. By using $J_{der1}(J_{der1-i})$ and $J_{der2}(J_{der2-i})$ as two individual objective functions, we obtain a set of Pareto points \textcolor{black}{\cite{PhysRevA.78.033414}} as displayed in Fig.\ref{Fig6:twoobject}, which explicitly shows an inverse relationship between them. As expected, the suppression of infidelity fluctuation (vertical axis) to experimental imperfections requires a larger sacrifice in the ideal fidelity (horizontal axis). E.g. for the DER$_{0.8}$ pulse, if we require the ideal infidelity close to $10^{-6}$(point 1) the fidelity fluctuations have to stay at a very high level $\sim 1.2\times 10^{-2}$ indicating a poor robustness to the detuning errors. But, when we could endure a relatively high  gate infidelity $1-\mathcal{F}_{der} \approx 2.8\times 10^{-3}$ (point 4), then the insensitivity can be lowered to $\sim 5.4\times 10^{-3}$.

\textcolor{black}{However, depending on a fine optimization, the DER-i$_{0.3}$ pulse is able to achieve a higher fidelity at $\epsilon_\delta=0$, while still keeping strong robustness to the detuning errors. E.g. here if $1-\mathcal{F}_{der-i} \approx 1.2\times 10^{-5}$ (point 1) the average insensitivity stays at $1.1\times 10^{-4}$. In contrast if $1-\mathcal{F}_{der-i} \approx 3.6\times 10^{-5}$ (point 4) the insensitivity to detuning errors can be as small as $8.9\times 10^{-5}$. Hence, due to the competing effect the optimal pulses used for implementing the gate should satisfy a trade-off between two aspects: ideal fidelity and robustness. However, we note that the fine optimization procedure is able to provide greatly-improved gate performance maintaining a high gate fidelity as well as {\it actively} strong robustness against the detuning errors.}

\section{Passive influences by other noises}

In order to characterize the gate performance in the presence of all experimental imperfections, in this section, we study other errors owing to the detuning deviation from the intermediate state $|p\rangle$, inhomogeneity in Rabi frequency, laser phase and amplitude noises, as well as the Rydberg interaction fluctuations. We address that 
our robust gate is passively influenced by those errors so the DER and DER-i pulses do not have explicit robustness to them as compared to the TO pulse.

\begin{figure}
    \centering
    \includegraphics[scale=0.33]{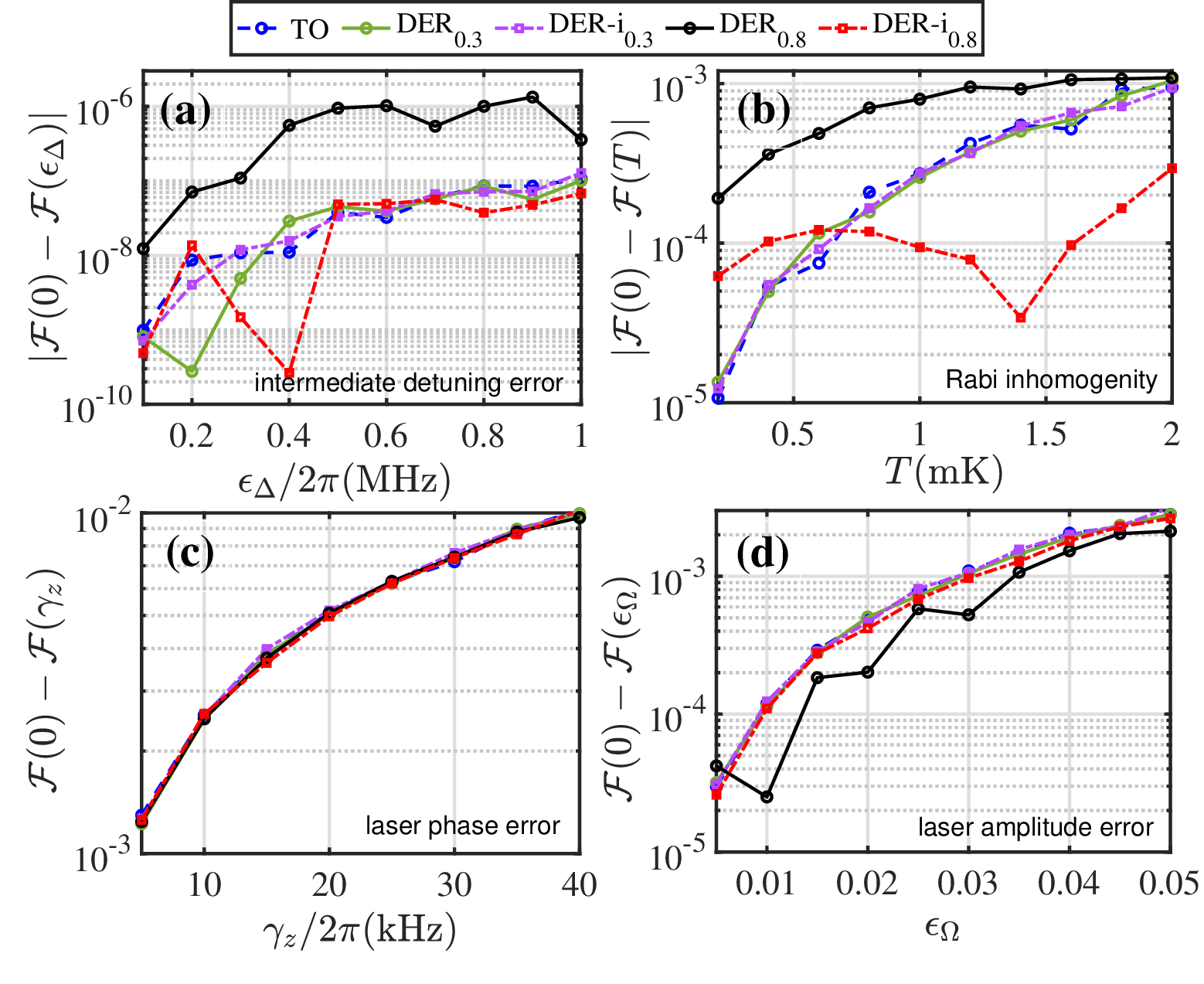}
    \caption{The gate infidelity of the TO, DER and DER-i (Gaussian-weighted) pulses caused by other error sources including (a) The detuning error due to the intermediate detuning deviation $\epsilon_\Delta$(note that the non-fluctuated part is $\Delta_0/2\pi=2.0$ GHz); (b) Inhomogeneous laser Rabi frequency coming from atomic position fluctuations associated with the variation of temperature $T$; (c) Average dephasing effect denoted by rate $\gamma_z$ due to the time-dependent laser phase noise; (d) Laser amplitude error caused by the amplitude deviation $\epsilon_{\Omega}$. Each point in all figures is obtained by averaging over 500 independent realizations. }
    \label{Fig7:allerror}
\end{figure}

\textit{Intermediate detuning error -} To assess the performance of gate in a fluctuated intermediate-state detuning $\Delta_0\to\Delta_0+\epsilon_\Delta$ (also see Eq.\ref{Hj}), we now study the gate infidelity with a randomly distributed error $\epsilon_\Delta$.
In a realistic experiment, this intermediate detuning error is inevitable due to the laser frequency fluctuation or atomic temperature effect \textcolor{black}{\cite{PhysRevA.80.013417}}. However, since a large non-fluctuated intermediate detuning $|\Delta_0|\gg\Omega_{p,c}$ is set for a two-photon process; therefore,  $\epsilon_\Delta\ll|\Delta_0|$ which could make this detuning error negligible.

We numerically verify the negligible effect of $\epsilon_\Delta$ on the absolute gate infidelity 
$|\mathcal{F}(0)-\mathcal{F}( \epsilon_\Delta)|$ for all pulses as shown in Fig.\ref{Fig7:allerror}(a). Note that the fluctuated detuning $\Delta_0+\epsilon_\Delta$ is also reversed to $-(\Delta_0+\epsilon_\Delta)$ halfway for cancelling the dynamical phase. It is clear that the infidelity caused by different $\epsilon_\Delta$ values stays below $10^{-6}$. This is expected because the large intermediate detuning $\Delta_0$ can efficiently suppress the intermediate population during the gate duration leading to a negligible error \textcolor{black}{\cite{QuantumSci.Technol.7(2022)045020}}. In addition we observe all pulses except DER$_{0.8}$ reveal comparably better robustness $< 10^{-7}$ to the $\epsilon_\Delta$ because of the presence of dark state $|d\rangle$(see Sec. III.A).
The DER$_{0.8}$ suffers from a slightly larger infidelity so the intermediate detuning error turns to be the worst.
While the DER and DER-i pulses have shown their improved robustness against the deviation of $\epsilon_\delta$, not necessarily the most robust for the $\epsilon_\Delta$. Finally since $\Delta_0$ is not an optimal value, we also observe the infidelity oscillations with respect to the change of $\epsilon_\Delta$.

\textit{Inhomogeneous Rabi frequency -} 
In general the position fluctuations of atoms are inevitable in experiments, which directly leads to a position-dependent laser Rabi frequency. Therefore the realistic Rabi frequency perceived by the atoms will deviate from its desired value. According to Ref.\cite{PhysRevA.85.042310}, we replace $\Omega_p(t)$ and $\Omega_c(t)$ by position-dependent functions $\Omega_p(\bold{r},t)$ and $\Omega_c(\bold{r},t)$, described by
\begin{equation}
\begin{aligned}
\Omega_{p}(\bold{r},t)=\Omega_{p}(0,t)\frac{e^{-\frac{x^{2}}{\omega_{x,p}^{2}(1+\frac{z^{2}}{L_{x,p}^{2}})}-\frac{y^{2}}{\omega_{y,p}^{2}(1+\frac{z^{2}}{L_{y,p}^{2}})}}}{[(1+\frac{z^{2}}{L_{x,p}^{2}})(1+\frac{z^{2}}{L_{y,p}^{2}})]^{1/4}}
\end{aligned}
\end{equation}
\begin{equation}
\begin{aligned}
\Omega_{c}(\bold{r},t)=\Omega_{c}(0,t)\frac{e^{-\frac{x^{2}}{\omega_{x,c}^{2}(1+\frac{z^{2}}{L_{x,c}^{2}})}-\frac{y^{2}}{\omega_{y,c}^{2}(1+\frac{z^{2}}{L_{y,c}^{2}})}}}{[(1+\frac{z^{2}}{L_{x,c}^{2}})(1+\frac{z^{2}}{L_{y,c}^{2}})]^{1/4}}
\end{aligned}
\end{equation}
where $\Omega_{p,c}(0,t) = \Omega_{p,c}(t)$ is the originally optimized waveform (see Eqs. \ref{opa}-\ref{opb}) and $\bold{r} = (x,y,z)$ presents the atom position at any time $t$ that satisfies a Gaussian distribution
\begin{equation}
\begin{aligned}
f(\bold{r}) = \frac{1}{(2\pi)^{3/2}\sigma_{x}\sigma_{y}\sigma_{z}}e^{-\frac{x^{2}}{2\sigma_{x}^{2}}}e^{-\frac{y^{2}}{2\sigma_{y}^{2}}}e^{-\frac{z^{2}}{2\sigma_{z}^{2}}}
\end{aligned}
\end{equation}

When the laser waists are $\omega_{x,p}=\omega_{y,p}=7.8$ $\mu$m, $\omega_{x,c}=\omega_{y,c}=8.3$ $\mu$m, arising the Rayleigh lengths $L_{x(y),p}=455.08$ $\mu$m, $L_{x(y),c}=213.65$ $\mu$m, respectively for the 420 nm and 1013 nm lasers. The standard deviation $\sigma_{x,y,z}$ in $f(\bold{r})$ can be estimated by $\sigma_{x,y,z}=\sqrt{k_BT/m\omega_{x,y,z}^{2}}$ with trap frequencies $ \omega_{x,y,z}/2\pi = (147,117,35)$ kHz chosen from \cite{Naturephotonics16.724(2022)}, which arises the maximal uncertainty in the atom position $(\sigma_x,\sigma_y,\sigma_z) \approx (0.47,0.60,1.99)$ $\mu$m at $T=2$ mK. That means the realistic deviation of Rabi frequency felt by atoms is still very small owing to $\sigma_x\ll\omega_{x,p(c)}$, $\sigma_y\ll\omega_{y,p(c)}$ even at a high temperature.

Our numerical results are summarized in Fig.\ref{Fig7:allerror}(b), where the infidelity caused by inhomogeneous Rabi frequency varies with the atomic temperature $T$. We observe that, although the waist of a Gaussian laser can be adjustable experimentally by using different incident angles, for a reasonable temperature, atoms in the optical traps can still be uniformly illuminated by both Rydberg excitation lasers \textcolor{black}{\cite{PhysRevX.5.031015}}. Because the maximal deviation of atom position is smaller than the beam waist by orders of magnitude. From Fig.\ref{Fig7:allerror}(b) the robust pulses do not have explicit advantages against this inhomogeneity error, and the finely-optimal DER, DER-i pulses reveal an exponential increase in the gate infidelity, analogously to the TO pulse.
 This is because the originally optimized pulse $\Omega_{p,c}(0,t)$ is explicitly robust to the two-detuning error, but not the best to other errors such as the
 Rabi inhomogenity. Slightly shifting Rabi frequency via atom position could instead improve the gate insensitivity to the variation of temperature, {\it e.g.} for the DER-i pulse under a wide optimization.

\textit{Noises from excitation lasers - } The Rydberg excitation lasers will not only contribute significant phase noise, but also have amplitude fluctuations. The former could change the realistic Rabi frequency by adding phase terms as
\begin{equation}
\begin{aligned}
\Omega_{p,c}(t)\rightarrow\Omega_{p,c}(t)e^{i\phi_{p,c}(t)}
\end{aligned}
\end{equation}
due to different frequencies involved
in the excitation lasers. Here, $\phi_{p,c}(t)$ presents a random phase typically featuring time-dependent fluctuations \cite{PhysRevA.97.053803}.

Accounting for the fact that the laser phase noise has shown its contribution to the decay of ground-Rydberg Rabi oscillations, it is reasonable to utilize a global dephasing model to
quantify
the average effect of laser phase noise as done by \textcolor{black}{\cite{PhysRevA.101.043421}}. So we newly introduce a Liouvillian dissipative operator ${\mathcal{\hat{L}}}_{z}[\hat{\rho}]$, serving as an extra non-Hermitian term in the master equation (\ref{rho})
\begin{equation}
    {\hat{\mathcal{L}}_z}[{\hat{\rho}}] = \sum_{l\in\{c,t\}}\sum_{i \in\{1,2\}}{ {\hat{\mathcal{L}}}^{l}}_{i}{\hat{\rho}} {{\hat{\mathcal{L}}}_{i}^{l^{\dagger}}}-\frac{1}{2}\{{ {\hat{\mathcal{L}}}_{i}^{l^{\dagger}}}{ {\hat{\mathcal{L}}}^{l}}_{i},{\hat{\rho}}\}
\end{equation}
where 
\begin{eqnarray}
\hat{{\mathcal{L}}}_{1}^{l}&=& \sqrt{\gamma_{1}/2}(|p \rangle \langle p|-|1 \rangle \langle 1|)\nonumber\\
\hat{{\mathcal{L}}}_{2}^{l}&=& \sqrt{\gamma_{2}/2}(|r \rangle \langle r|-|p \rangle \langle p|)\nonumber
\end{eqnarray}
represent dephasings of the control and target atoms with individual rates $\gamma_1,\gamma_2$ associated with the optical transitions of $|p\rangle\to|1\rangle$ and $|r\rangle\to|p\rangle$, which are
randomly obtained from the range of $[0, \gamma_{z}] $ ($\gamma_z$ treats as the maximal dephasing rate). Figure \ref{Fig7:allerror}(c) shows the gate infidelity caused by the increase of $\gamma_z$. Due the impact of laser phase noise we see all pulses (including non-robust and robust pulses) exhibit a same trend with an exponential enhancement in the gate infidelity $\mathcal{F}(0)-\mathcal{F}(\gamma_z)$. When $\gamma_{z}/2\pi = 40$ kHz we find the average infidelity even reaches $\sim 10^{-2}$ after running 500 simulations.

Furthermore, the latter can be described by 
an uncertainty in Rabi frequencies
\begin{equation}
 \Omega_{p}(t)\rightarrow\Omega_{p}(t)(1+ \epsilon_{\Omega_{p}}),\Omega_{c}(t)\rightarrow\Omega_{c}(t)(1+ \epsilon_{\Omega_{c}})
 \label{ome}
\end{equation}
\textcolor{black}{To distinguish from the ac Stark shift error which is also induced by the amplitude deviation (see Sec. V.B.), here we consider the detuning error $\epsilon_\delta = 0$ and modify the sing-qubit Hamiltonian (\ref{Hj}) as
\begin{eqnarray}
    {\hat{H}_{j\in(c,t)}} 
    &=&\frac{\Omega_p(1+\epsilon_{\Omega_p})}{2}|p\rangle_j \langle 1|+\frac{\Omega_c(1+\epsilon_{\Omega_c})}{2}|r\rangle_j \langle p|+\text{H.c.} \nonumber \\
    & -&\Delta_0|p\rangle_j \langle p| \label{Hjn}
\end{eqnarray}
Assuming the amplitude deviations $\epsilon_{\Omega_p},\epsilon_{\Omega_c}$ are random within the range of $ [-\epsilon_\Omega,+\epsilon_\Omega]$ we calculate the infidelity $\mathcal{F}(0)-\mathcal{F}(\epsilon_\Omega)$ as a function of the amplitude error $\epsilon_\Omega$ in Fig.\ref{Fig7:allerror}(d). It is clearly seen that the laser amplitude noise plays a less significant impact as compared to the phase noise (the former), maintaining smaller than $2.5\times 10^{-3}$ for large values of deviation $\epsilon_\Omega$ up to 0.05, which is even regardless of the pulse types.}

\textit{Fluctuation of Rydberg interaction - }
Our scheme relies on the Rydberg blockade mechanism that are robust to the fluctuation of interatomic interaction \textcolor{black}{\cite{PhysRevApplied.7.064017}}. Under finite temperature, the atom position in a trap will lead to a fluctuation $\Delta B$ in the interaction strength deviating from its expected value $B$. 
While benefiting from a strong long-range interaction to suppress the excitation of $|rr\rangle$, we find the impact of interaction fluctuation is very small. 
A rough estimation shows that, if $\Delta B/B = 0.1$($B/2\pi=2.0$ GHz), the relative position deviation between two atoms is about $\delta r=  r_0^7\Delta B/(6C_6) \approx45.83$ nm \textcolor{black}{\cite{Chin.Phys.B32.083302(2023)}}, estimated by the position-dependent interaction strength $B = C_6/r_0^6$ with $C_6 =862.69$ $\rm{GHz}\cdot\mu m^6$($70S_{1/2}$) and $r_0 = 2.75$ $\mu$m \textcolor{black}{\cite{ARC}}.
Then the gate infidelity of all pulses stays below $10^{-10}$(not shown) providing a 
a negligible impact on the gate operations.

To briefly conclude, our proposed gates are passively affected by these error sources irrelated to the two-photon detuning, among which the laser phase noise plays the most dominant role. Because the use of adiabatic pulses will cause a prolonged gate duration, increasing the damping of Rabi oscillations for the ground-Rydberg transition. To reduce this phase noise, we may increase the peak amplitude that restricts the timescale for the gate duration \textcolor{black}{\cite{PhysRevA.90.032329}}, utilize the adiabatic rapid passage approach with single optimized pulse for a direct single-photon transition \textcolor{black}{\cite{J.Phys.B53(2020).182001,PhysRevA.101.030301}} or the superadiabatic protocols \textcolor{black}{\cite{Sci.Adv.5.eaau5999(2019)}}. In experiment, low-noise laser sources or diode lasers filtered by higher-finesse optical cavities will further eliminate the laser phase error \cite{PhysRevLett.121.123603,PhysRevApplied.15.054020}.

\nocite{*}
\bibliographystyle{apsrev4-2}

\end{document}